\documentclass[aps,amsmath,amssymb,showpacs,showkeys]{revtex4}
\usepackage[dvips]{graphicx,color}
\usepackage{times}
\usepackage{xcolor}
\usepackage[%
  colorlinks=true,
  urlcolor=blue,
  linkcolor=red,
  citecolor=blue
]{hyperref}
\usepackage{orcidlink}
\usepackage{mathrsfs}
\usepackage{amsmath}
\usepackage{accents}
\usepackage{mathtools}
\usepackage{amssymb}
\usepackage{graphicx}
\usepackage{epsfig}
\usepackage{dcolumn}
\usepackage{bm}
\usepackage{tikz}
\usetikzlibrary{arrows, decorations.markings, calc, fadings, decorations.pathreplacing, patterns, decorations.pathmorphing, positioning, angles, quotes}
\newcommand{\orcid}[1]{\href{https://orcid.org/#1}{\includegraphics[width=8pt]{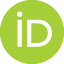}}}

\begin{document}
\title{Deflection angle, quasinormal modes and optical properties of a de 
Sitter black hole in $\bm{f(\mathcal{T}, \mathcal{B})}$ gravity}

\author{Nashiba Parbin \orcidlink{0000-0002-3570-0117}}
\email[Email: ]{nashibaparbin91@gmail.com}

\affiliation{Department of Physics, Dibrugarh University,
Dibrugarh 786004, Assam, India}

\author{Dhruba Jyoti Gogoi \orcidlink{0000-0002-4776-8506}}
\email[Email: ]{moloydhruba@yahoo.in}

\affiliation{Department of Physics, Dibrugarh University,
Dibrugarh 786004, Assam, India}

\author{Jyatsnasree Bora \orcidlink{0000-0001-9751-5614}}
\email[Email: ]{jyatnasree.borah@gmail.com}

\affiliation{Department of Physics, Dibrugarh University,
Dibrugarh 786004, Assam, India}

\author{Umananda Dev Goswami \orcidlink{0000-0003-0012-7549}}
\email[Email: ]{umananda2@gmail.com}

\affiliation{Department of Physics, Dibrugarh University,
Dibrugarh 786004, Assam, India}

%\date{}
\begin{abstract}
The current study aims to examine the impact of the boundary term on the 
bending angle of light for a static spherically symmetric black hole in
the modified gravity described by the $f(\mathcal{T}, \mathcal{B})$ function. 
To accomplish this objective, we employ the Ishihara \textit{et al.}~method, 
which enables us to compute the deflection angle of light for a receiver and 
source situated at finite distances from a lens object in a non-asymptotically 
flat spacetime. This method considers the receiver's viewpoint, and the 
resulting deflection angle diverges as the distance from the lens object 
increases, owing to the non-asymptotically flat spacetime. Nevertheless, the 
divergence can be regulated by the boundary term parameter $c_0$. For lower 
values of the parameter $c_0$, the divergence can be minimized within the 
finite range of the observer and source. Furthermore, we calculate the 
quasinormal modes of massless scalar perturbations in the black hole's 
background using the asymptotic iteration method (AIM) and Pad\'e averaged 
sixth-order Wentzel-Kramers-Brillouin (WKB) approximation method. Our findings 
indicate that the real quasinormal modes and damping rates are significantly 
impacted by the model parameter $c_0$. Subsequently, we investigate two 
optical characteristics of the black hole, namely the shadow and the emission 
rate. Our results show that with an increase in the boundary term parameter 
$c_0$, both the shadow's size and the evaporation rate decrease.
\end{abstract}

%\pacs{04.30.Tv, 04.50.Kd}

\maketitle

\section{Introduction}
\label{sec.1}

The revolutionary idea of the theory of general relativity (GR) first struck 
in the great mind of Albert Einstein, which he published in $1915$ 
\cite{einstein}. After the time of publication of this theory, it has gained 
utmost importance in describing many gravitational phenomena. Among different 
predictions of GR, black holes and gravitational waves (GWs) are two most 
significant ones. However, it has shown some limitations while explaining the 
current accelerated expansion of the universe \cite{reiss, perlmutter} as well
as the observed rotational dynamics of the galaxies \cite{rubin, young1, harko, gergely, parbin}. GR needs some exotic stuff called dark energy and dark matter 
\cite{bertone, swart, tim, frenk, strigari,will2014} in order to overcome its 
limitations. Thus, in order to represent the observed facts of the present 
universe, it can be conjectured 
that either the universe is filled with mysterious dark matter and dark 
energy, or a plausible modified theory of gravity (MTG) should be used such 
that it can explain the correct outcomes from observations. At present there 
are lots of proposed MTGs that are introduced to resolve unexplained 
issues of the universe \cite{Clifton, wagoner, ronald, capozziello, clifton, 
oikonomou, odintsov, sergei, gogoi1, nashiba, gogoi7,gogoi_cosmo}. 

The first experimental verification of GR was observed in $1919$. It was the 
observation of the gravitational bending of light, during a solar eclipse 
\cite{eddington} as per the prediction of GR. Moreover, from the 
interpretation of gravitational bending of light in the framework of 
geometrical optics of a lens, a key idea, the gravitational lensing has 
emerged \cite{schneider, trimble, renn, valls}. It was in $1979$, when the 
first instance of gravitationally lensed object, the double quasar 
Q$0957 + 561$A,B was identified for the first time \cite{young}. In the modern 
observational cosmology for exploring the exoplanets and for measuring the 
distribution of dark matter and dark energy, an important parameter is the angle 
of gravitational bending of light. For different black holes, the gravitational 
lensing around them has been investigated in a plethora of articles. 
For the case of Schwarzschild black hole such investigation was carried out in 
the Ref.\,\cite{virbhadra}. For black holes in presence of the cosmological 
constant the studies of such lensing are reported in 
Ref.s \cite{zhao,rindler}. Again, the gravitational lensing by the naked 
singularity and horizonless ultracompact objects are studied in the 
Ref.s \cite{ellis, shaikh}.

In 2008, an alternative method for the derivation of deflection angle of light 
in a spherically symmetric black hole, introduced by Gibbons and Werner, by 
implementing the Gauss-Bonnet theorem (GBT) \cite{carmo, bonnet} was reported 
in the Ref.\,\cite{gibbons}. This method is found to be good to calculate the 
exact form of the deflection angle in the weak field limit for Schwarzschild 
black hole spacetime \cite{gibbons}. Quite a few studies show that this 
approach can be used to derive the deflection angle for different black hole 
solutions \cite{banerjee, jusufi, kimet, sakalli, Kjusufi, izzet, ovgun1, ovgun2, ovgun3, anisur, eslam, panah}. To obtain the deflection angle by a Kerr black hole the 
Gibbons-Werner method was then extended using the Kerr-Randers optical 
geometry \cite{werner, saavedra}. This method was further extended in $2016$ 
by Ishihara $\textit{et al.}$ to consider finite distances between the source 
and the observer for static as well as stationary black hole solutions 
\cite{ishihara}. For stationary black hole spacetimes, extension of this 
method is reported in the Ref.s \cite{ono,kumar,ghosh, jamil,crisnejo}. For 
the case of asymptotically non-flat spacetimes such study can be found in the 
Ref.s \cite{asada, takizawa, carvalho}. Using different MTGs such studies on 
gravitational deflection angle using GBT are reported in 
\cite{carvalho, jamil, mubasher}. Using GBT the effect of dark matter on 
deflection angle is also reported earlier in a few articles 
\cite{pantig1, pantig2, javed, ovgun4}.

In our work, we intend to explore the impact of the modification of spacetime 
curvature on the gravitational lensing of a static spherically symmetric black 
hole spacetime. For this purpose, we consider the black hole solution in the 
$f(\mathcal{T},\mathcal{B})$ modified gravity theory \cite{2015_Bahamonde}. 
The $f(\mathcal{T},\mathcal{B})$ modified gravity has been investigated in 
various aspects \cite{2017_Bahamonde,2018_Bahamonde,2020_cappozielo,2020_farrugia,2022_Bahamonde, camci}, but for the case of gravitational lensing, we 
study for the first time the effects of $f(\mathcal{T},\mathcal{B})$ gravity 
on the deflection angle of light for a static spherically symmetric black hole 
using the GBT. Here we shall implement the Ishihara method for the 
asymptotically non-flat spacetime to obtain the deflection angle.

In this work we also study the quasinormal modes \cite{rayimbaev, ghasemi} of 
the considered black hole spacetime. In general these are the modes of 
emission of GWs from compact and massive perturbed objects in the universe and 
are represented by some complex numbers \cite{1970_Vishveshwara,1971_press,1975_Chandrasekhar, chandrasekhar}. The real part of these complex numbers is 
related to the emission frequency, while the imaginary part is related to the 
damping of quasinormal modes of GWs. In the last few years a number of authors 
have investigated the properties of GWs and also quasinormal modes of black 
holes in different MTGs \cite{2008_ca,gogoi1,2017_liang,2021_oliveira,2021_gogoi,2018_graca,2022_gogoi, gogoi2, gogoi4, gogoi5, gogoi6}.

Furthermore, we analyze the effect that the modification of curvature has on 
the shadow cast by the black hole in $f(\mathcal{T},\mathcal{B})$ gravity. The 
shadow is a characteristic feature of a black hole and is cast as a result of 
the strong gravitational field of the black hole. The shape and size of the 
shadow depend directly on the mass and angular momentum of the black hole. 
Studies on the shadow cast by a black hole have been of current interest to the 
scientific community \cite{ovgun5, wei, cunha, wang, dastan, jamil, kumar,1s,shnew02,shnew01,2s,3s,4s,5s,6s,7s,8s,9s,10s,11s,12s,13s,14s,15s, 16s,17s,18s,19s,102-1,102-2,102-3,102-4,102-5,102-6,102-7,102-8,102-9,102-10,102-11,
102-12,102-13,102-14,102-15}, more so after the successful release of the black hole images. The studies of the shadow cast by black holes surrounded by exotic 
dark matter have also been carried out in various literature \cite{haroon, salucci, hou, konoplya}. Also, the analysis of the shadow cast by a black hole can 
aid in characterizing the various gravity theories \cite{3s}. Finally, we 
discuss and analyze the behaviour of the emission rate of particles around the 
black hole. It is another observable phenomena which has also gained interest 
and has been studied in literature \cite{Wei2013, mashhoon, page, crispino, higuchi, oliviera, carlos}.

Our paper is organized as follows. In Sec.\ \ref{sec.2}, we briefly review the 
field equations related to the $f(\mathcal{T},\mathcal{B})$ modified gravity 
theory. In Sec.\ \ref{sec.3}, we derive the deflection angle in the spacetime 
of the static black hole in $f(\mathcal{T},\mathcal{B})$ gravity applying 
Ishihara $\textit{et al.}$ method for asymptotically non-flat spacetime. The 
quasinormal modes of the black hole are studied in Sec.\ \ref{sec.4}. In 
Sec.\ \ref{sec.5}, we have studied the optical properties, namely, the shadow 
and emission rate of the black hole. Finally, in Sec.\ \ref{sec.6}, 
we summarize and conclude the results of our work. Throughout our work, we 
have imposed the sign convention ($-, +, +, +$).

\section{$f(\mathcal{T}, \mathcal{B})$ gravity and a static spherically 
symmetric black hole solution} 
\label{sec.2}
An alternative geometrical formulation of GR can be made through the 
teleparallel equivalent of general relativity (TEGR). In other words the TEGR 
is an analogous theory of GR as this theory asserts that GR can be formulated
from the tetrad fields as well as from the torsion tensor 
\cite{2013_Aldrovandi,2013_Maluf}. A generalization of the TEGR is the 
$f(\mathcal{T})$ gravity, a function of the torsion scalar 
$\mathcal{T}$, which is obtained by the contraction of the torsion 
tensor ${\mathcal{T}^a}_{\mu\nu}$ defined as \cite{2015_Bahamonde,2013_Maluf}
\begin{equation}
{\mathcal{T}^a}_{\mu\nu} = \partial_\mu {h^a}_\nu - \partial_\nu {h^a}_{\mu},
\label{eqn.a}
\end{equation}
where ${h^a}_\mu$ are the tetrad fields, related to the metric $g_{\mu\nu}$ 
via the Minkowski metric $\eta_{ab}$ as
\begin{equation}
g_{\mu\nu} = {h^a}_{\mu}{h^b}_{\nu}\eta_{ab}. 
\label{eqn.b}
\end{equation}
A similar relationship between the inverse metric $g^{\mu\nu}$ and the inverse 
tetrads ${h_a}^\mu$ can be obtained. The conditions satisfied by the tetrads 
and the inverse tetrads are
\begin{equation}
{h_b}^\mu {h^a}_\mu = \delta^a_b, \;\;\;\; {h_a}^\mu {h^a}_\nu = 
\delta^\mu_\nu. 
\label{eqn.c}
\end{equation}    
A more generalization to the $f(\mathcal{T})$ gravity is done by introducing a 
boundary term $\mathcal{B}$ related to the divergence of the torsion 
vector $\mathcal{T}_\mu = {\mathcal{T}^\alpha}_{\alpha\mu}$ as 
$\mathcal{B}=\frac{2}{h}\,\partial_\mu(h\,\mathcal{T}^\mu)$, where 
$h=det({h^a}_\mu)=\sqrt{-g}$ is the determinant of the tetrads ${h^a}_\mu$ 
with $g$ being determinant of the metric $g_{\mu\nu}$. This theory was 
introduced in \cite{2015_Bahamonde}. The action of the 
$f(\mathcal{T}, \mathcal{B})$ gravity is
\begin{equation}
S_{f(\mathcal{T}, \mathcal{B})}=\int d^4x\,h\!\left[\frac{1}{2\kappa^2} f(\mathcal{T}, \mathcal{B}) + \mathcal{L}_m \right],
\label{eqn.1}
\end{equation}
with $\kappa^2=8\pi G$ and $\mathcal{L}_m$ being the Lagrangian density of 
matter. 
%The term $h$ is the tetrad determinant and is given by 
%\begin{equation}
%h=det({h^a}_{ \mu})=\sqrt{-g}.
%\label{eqn.2}
%\end{equation}
The field equations of this theory can be obtained as 
\cite{2015_Bahamonde,2022_Bahamonde}
\begin{eqnarray}\nonumber
\delta^\lambda_\nu\accentset{\scriptstyle \circ}{\square} f_\mathcal{B}-\accentset{\scriptstyle \circ}{\nabla}^\lambda\accentset{\scriptstyle \circ}{\nabla}_\nu f_\mathcal{B} + \frac{1}{2}\mathcal{B} f_\mathcal{B}\delta^\lambda_\nu+\Big[(\partial_\mu f_\mathcal{B})+(\partial_\mu f_\mathcal{T})  
\Big]{S_\nu}^{\mu\lambda} 
+h^{-1}{h^a}_\nu\partial_\mu\!\left(h{S_a}^{\mu\lambda}\right)\!f_\mathcal{T}\\
\quad\quad\quad\quad\quad\quad 
-f_\mathcal{T} {\mathcal{T}^\sigma}_{\mu\nu}{S_\sigma}^{\lambda\mu}-\frac{1}{2}f\delta^\lambda_\nu = \kappa^2T_\nu^\lambda.
\label{eqn.3}
\end{eqnarray}
In this expression the circle over the terms is used to denote the quantities
which are determined using the Levi-Civita connection and the subscripts below 
the functional $f$ denote the respective derivatives. The term $T_\nu^\lambda$ 
represents the energy-momentum tensor and the term ${S_\nu}^{\mu\lambda}$ is 
known as the superpotential tensor \cite{2022_Bahamonde} defined as
\begin{equation}
{S_\nu}^{\mu\lambda}={K^{\mu\lambda}}_{\nu}-\delta^\mu_\nu {\mathcal{T}_\sigma}^{\sigma\lambda}+\delta^\lambda_\nu {\mathcal{T}_\sigma}^{\,\sigma\mu}=-{S_\nu}^{\lambda\mu},
\end{equation}
where ${{K^\nu}}_{\mu\lambda}$ is the contortion tensor as given by
\begin{equation}
{K^\nu}_{\mu\lambda}={\Gamma^\nu}_{\mu\lambda}-{\accentset{\scriptstyle \circ}{{\Gamma}}^\nu}_{\mu\lambda}=\dfrac{1}{2}\left({{\mathcal{T}_\mu}^\nu}_\lambda+{{\mathcal{T}_\lambda}^\nu}_\mu-{\mathcal{T}^\nu}_{\mu\lambda}  \right).
\end{equation} 

In Ref.~\cite{2022_Bahamonde} different exact and spherically symmetric 
perturbative solutions of field Eqs.~\eqref{eqn.3} have been obtained. 
The black hole solution that we consider in our work is obtained 
by using the complex tetrad found in the Weitzenb\"ock gauge (${\omega^A}_{B\mu} = 0$) \cite{krssak}. 
This complex tetrad has been uniquely derived in \cite{2022_Bahamonde}, and is given by
\begin{equation}
h^A_{(2)\mu} = \begin{pmatrix}
0 & iB(r) & 0 & 0 \\
iA(r)\sin\vartheta\cos\phi & 0 & -\chi r \sin\phi & -r \chi \sin\vartheta\cos\vartheta\cos\phi \\
iA(r)\sin\vartheta\cos\phi & 0 & \chi r \cos\phi & -r \chi \sin\vartheta\cos\vartheta\sin\phi \\
iA(r)\cos\vartheta & 0 & 0 & \chi r \sin^2\vartheta
\end{pmatrix}, \hspace{3mm} (\chi = \pm 1),
\label{eqn.3a}
\end{equation}
where, $A(r)$ and $B(r)$ are the metric elements for a 
spherically symmetric metric of the form
\begin{equation}
ds^2 = - A(r)\, dt^2 + B(r)dr^2 + r^2\, d\Omega^2,
\end{equation}
where $d\Omega^2 \equiv d\theta^2 + \sin^2\!\theta\, d\phi^2$. 

Following the approach described in \cite{2022_Bahamonde} and 
using the complex tetrad (\ref{eqn.3a}) in the field equations (\ref{eqn.3}) 
of $f(\mathcal{T}, \mathcal{B})$ gravity theory, the symmetric field equations 
for the complex tetrad are obtained as
\begin{equation}
\begin{split}
\kappa^2 \rho &= -\frac{1}{2}f + \frac{2f_\mathcal{T} (rB A^\prime + A(B - rB^\prime))}{r^2 AB^3} + 
\frac{f_\mathcal{B}(r(B(rA^{\prime\prime} + 4A^\prime) - rA^\prime B^\prime) + 
2A(B - rB^\prime))}{r^2 AB^3} \\
& + \frac{B^\prime f_\mathcal{B}^\prime}{B^3} - \frac{f_\mathcal{B}^{\prime\prime}}{B^2} + 
\frac{2f_\mathcal{T}^\prime}{rB^2} ,
\end{split}
\label{eqn.3b}
\end{equation}
\begin{equation}
\begin{split}
\kappa^2 p_r &= \frac{1}{2}f + \frac{(rA^\prime + 2A)f_\mathcal{B}^\prime}{rAB^2} - 
\frac{2f_\mathcal{T}(2rA^\prime + A)}{r^2 A B^2} + \frac{f_\mathcal{B}(r(rA^\prime B^\prime - 
B(rA^{\prime\prime} + 4A^\prime))- 2A(B - rB^\prime))}{r^2 AB^3} ,
\end{split}
\label{eqn.3c}
\end{equation}
\begin{equation}
\begin{split}
\kappa^2 p_l &= \frac{1}{2}f - \frac{(rA^\prime + A)f_\mathcal{T}^\prime}{rAB^2} + 
\frac{f_\mathcal{B}(r(rA^\prime B^\prime - B(rA^{\prime\prime} + 4A^\prime))- 2A(B - rB^\prime))}{r^2 AB^3} \\
&+ \frac{f_\mathcal{T}(r(rA^\prime B^\prime - B(rA^{\prime\prime} + 3A^\prime))- A(B - rB^\prime + B^3))}{r^2 AB^3} 
- \frac{B^\prime f_\mathcal{B}^\prime}{B^3} + \frac{f_\mathcal{B}^{\prime\prime}}{B^2},
\end{split}
\label{eqn.3d}
\end{equation}
where, the primes represent derivation with respect to the radial coordinate $r$, and 
$\rho$, $p_r$ and $p_l$ are the energy density, radial pressure and lateral pressure of the 
fluid respectively.

Again, as reported in \cite{2022_Bahamonde}, for a functional form of the type 
$f(\mathcal{T},\mathcal{B}) = \kappa_1 \mathcal{T} + F(\mathcal{B})$, 
Eqs.\ \eqref{eqn.3b}, \eqref{eqn.3c} and \eqref{eqn.3d} were utilized to 
obtain
\begin{equation}
- \frac{\kappa_1 (r(B(rA^{\prime\prime} + A^\prime) - rA^\prime B^\prime) + A(rB^\prime + B^3 - B))}{r^2 A B^3} = 0.
\label{eqn.3e}
\end{equation}
Solving this equation led to the derivation of an exact solution which 
behaves as Schwarzschild-de Sitter black hole solution. This black hole solution is 
of our interest as we wish to study the phenomenon of gravitational lensing by a black 
hole and its quasinormal modes in $f(\mathcal{T}, \mathcal{B})$ gravity theory. 
The Schwarzschild-de Sitter black hole solution is obtained by considering $\kappa_1 = 1/2$ 
and relating the metric elements as $A(r) = 1/B(r) = f(r)$ and is given by
\begin{equation}
ds^2 = -f(r)\, dt^2 + \frac{dr^2}{f(r)} + r^2\, d\Omega^2,
\label{eqn.4}
\end{equation}
where, the metric co-efficient is obtained as
\begin{equation}
f(r) = 1 - \frac{2M}{r} - (\Lambda + c_0 M)\,r^2. 
\label{eqn.4a}
\end{equation}

The functional form of the $f(\mathcal{T},\mathcal{B})$ gravity model 
considered in this solution is \cite{2022_Bahamonde}
\begin{equation}
f(\mathcal{T},\mathcal{B})=\frac{\mathcal{T}}{2}-\dfrac{8\,c_0}{3\sqrt{\mathcal{B}+18(Mc_0+\Lambda)}}-3\Lambda.
\label{eqn.5}
\end{equation}
This functional form of $f(\mathcal{T},\mathcal{B})$ is motivated from
the fact that it leads to the unique Schwarzschild-de Sitter solution
\eqref{eqn.4}. It can be obtained by choosing $\kappa_1 = 1/2$ in the 
generalized functional form $f(\mathcal{T,B}) = \kappa_1 \mathcal{T} + 
F(\mathcal{B})$, mentioned earlier, and deducing the remaining part of this 
relation, i.e. $F(\mathcal{B})$, by using Eq.~\eqref{eqn.3e} following the 
Ref.~\cite{2022_Bahamonde}. The  solution for the term $F(\mathcal{B})$ is 
made in such a way that the model presented in Eq.~\eqref{eqn.5} leads to the 
Schwarzschild-de Sitter solution~\eqref{eqn.4} (see \cite{2022_Bahamonde} for
details). Here $c_0$ is a parameter of the model carrying the effect of the 
boundary term $\mathcal{B}$. It is a dimensional parameter having the 
dimension of Length$^{\,-\,3}$. Thus the solution is found to have an 
effective cosmological constant due to the contribution from the boundary term 
as given by $\Lambda_{eff} = \Lambda + c_0 M$. One can estimate an 
observational limit on $c_0$ from the shadow of black holes. For example,
considering the black hole metric \eqref{eqn.4}, we can have an expression of 
the shadow of an asymptotically flat black hole for a static observer 
at large distance as given by (details are discussed in the Section \ref{sec.5}(A))
\begin{equation}
    R_{s} = \frac{3 M}{\sqrt{\frac{1}{3}-9 M^2 \left(c_0 M+\Lambda \right)}}.
\end{equation}
Following Ref. \cite{sh01} and considering M$87^*$ black hole shadow, it is possible to have the following cases,
\begin{equation}
    c_0 \le 1.11\times 10^{-49}- 2.15\times10^{-21} \Lambda \;\;\; \text{kpc}^{-3},
\end{equation}
and
\begin{equation}
     c_0 \le 1.59\times10^{-49}-2.15\times10^{-21} \Lambda \;\;\; \text{kpc}^{-3}.
\end{equation}

Although the tetrad used for the derivation of the exact black hole solution 
is complex, however the torsion scalar $\mathcal{T}$ and the boundary term 
$\mathcal{B}$ are real, and eventually, for the black hole solution \eqref{eqn.4}, the torsion scalar and the boundary term are obtained as
\begin{equation}
\mathcal{T} = \frac{4}{r^2} + (\Lambda + c_0 M)\left(\frac{16M}{r} - 12\right),
\end{equation}
\begin{equation}
\mathcal{B} = \frac{4}{r^2} + (\Lambda + c_0 M)\left(\frac{40M}{r} - 36\right).
\end{equation}

One important point to be noted here is that the black hole solution \eqref{eqn.4} 
is not asymptotically flat because of
the presence of $r^2$ term in the last part of the metric function 
\eqref{eqn.4a}. In this work we shall use this black hole solution to study 
the gravitational deflection of light and the quasinormal modes as discussed 
in the following sections. 

\section{Gravitational deflection of light}
\label{sec.3}

\begin{figure}[ht!]
\includegraphics[scale=0.35]{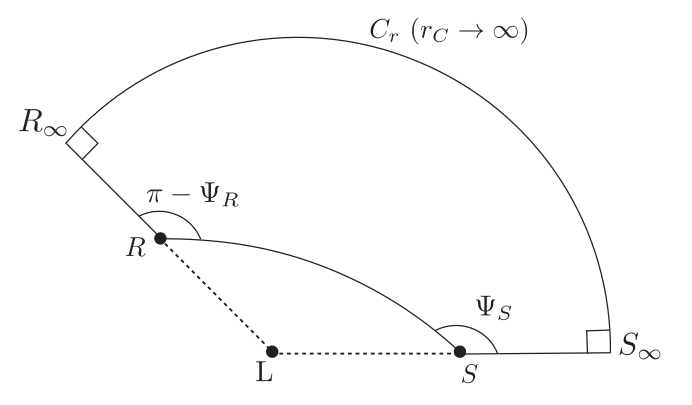}
\caption{Schematic figure for the quadrilateral $\stackrel{\infty}{R}\!\square\!\stackrel{\infty}{S}$ embedded in a curved space \cite{takizawa}.}
 \label{fig.1}
\end{figure}

We follow the Ishihara $\textit{et al.}$ approach \cite{ishihara} to obtain 
the deflection angle of light in the weak field limit of the non-asymptotically flat black hole spacetime given by the solution \eqref{eqn.4}. Here, as shown 
in Fig.\ \ref{fig.1}, the black hole works as a lens ($L$), which is at a 
finite distance from the source ($S$) and the receiver ($R$). In the 
equatorial plane ($\theta = \pi/2$), the deflection angle can be expressed as 
\cite{ishihara, ono}
\begin{equation}
\hat{\Theta} = \Psi_R - \Psi_S + \phi_{RS},
\label{eqn.6}
\end{equation}
where $\Psi_R$ and $\Psi_S$ are the angles of light that are measured with 
respect to the lens at the positions of the receiver and the source 
respectively. $\phi_{RS} = \phi_R - \phi_S$ is the separation angle between 
the receiver and the source. Here, $\phi_R$ and $\phi_S$ are the longitudes of 
the receiver and the source respectively. 

Since the light rays follow the null geodesic for which $ds^2 = 0$, the metric 
Eq.~(\ref{eqn.4}) can be rewritten for this geodesic as
\begin{equation}
dt^2 = \gamma_{ij} dx^i dx^j = \frac{1}{f(r)^2}\, dr^2 + \frac{r^2}{f(r)}\, d\Omega^2,
\label{eqn.7}
\end{equation}
where $\gamma_{ij}$ is usually known as the optical metric, which specifies a 
3D Riemannian space. We will denote this space by $\mathcal{M}^{(3)}$, where a 
ray of light is considered as a spatial curve. By using the optical metric 
$\gamma_{ij}$, we can define the angles $\Psi_R$ and $\Psi_S$. The 
non-vanishing components of this metric are
\begin{equation}
\gamma_{rr} = \frac{1}{f(r)^2},\;\; \gamma_{\phi\phi} = \frac{r^2}{f(r)}.
\label{eqn.8}
\end{equation}

An important parameter in the study of gravitational deflection angle of light
is the impact parameter of light in the black hole spacetime. This is usually
defined as the ratio of the angular momentum ($L$) to the energy ($\mathcal{E}$) of 
photons, which are the constant of motion in the equatorial plane of 
spacetime. For the spacetime of black hole \eqref{eqn.4} these two constants 
of motion are $\mathcal{E} = f(r)\,\dot{t}$ and $L = r^2\,\dot{\phi}$, where the over 
dot denotes the derivative with respect to the affine parameter $\lambda$ 
along the path of the light ray. Thus the impact parameter of a light ray is  
\begin{equation}
\xi \equiv \frac{L}{\mathcal{E}} = \frac{r^2}{f(r)} \frac{d\phi}{dt}.
\label{eqn.9}
\end{equation}
The unit radial vector from the center of the lens can be obtained as 
$e_{rad} = (f(r) , 0)$, and the unit angular vector along the angular direction 
can be found as $e_{ang} = (0, f(r)/r)$. Again, the components of the unit 
tangent vector $\bm{K} \equiv d\bm{x}/dt$ along the light ray are obtained 
as \cite{ishihara}
\begin{equation}
(K^r , K^\phi) = \frac{\xi f(r)}{r^2} \left(\frac{dr}{d\phi} , 1\right).
\label{eqn.10}
\end{equation}
Here $dr/d\phi$ gives the orbital path variation of the rays of light known 
as the orbit equation, which can be expressed as  
\begin{equation}
\left(\frac{dr}{d\phi}\right)^{\!2} = -\, r^2 f(r) + \frac{r^4}{\xi^2}.
\label{eqn.13}
\end{equation}
Now, if $\Psi$ denotes the angle between the radial component of the tangent 
vector and the radial vector, i.e.~the angle of the light ray which is 
measured from the radial direction, then we can write, 
\begin{equation}
\cos\Psi = \frac{\xi}{r^2} \frac{dr}{d\phi}.
\label{eqn.11}
\end{equation}
It gives,
\begin{equation}
\sin\Psi = \frac{\xi \sqrt{f(r)}}{r}.
\label{eqn.12}
\end{equation}
%Finally, we arrive at the orbit equation obtained as,
%\begin{equation}
%\left(\frac{dr}{d\phi}\right)^2 = - r^2 f(r) + \frac{r^4}{\xi^2 f(r)^2}
%\label{eqn.13}
%\end{equation}
Moreover, in a general form Eq.~\eqref{eqn.13} can be expressed as
\begin{equation}
\left(\frac{du}{d\phi}\right)^2 = F(u),
\label{eqn.14}
\end{equation}
where we have considered a new variable $u = 1/r$ and hence the function 
$F(u) = -\,u^2\,f(u) + 1/\xi^2$.

\begin{figure}[ht!]
\includegraphics[scale=0.55]{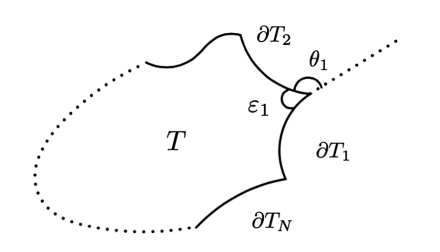}
\caption{Schematic figure for Gauss-Bonnet theorem \cite{ishihara}. The inner
angle is $\varepsilon_a$ and the jump angle is $\theta_a$ ($a =1,2, ...,N$).}
\label{fig.2}
\end{figure}

As mentioned earlier, we will use the GBT to calculate the deflection angle 
$\hat{\Theta}$. The GBT has many formulations. The simplest one states that the
total Gaussian curvature of an embedded triangle can be expressed in terms of
the total geodesic curvature of the boundary and the jump angles at the 
corners. Thus, mathematically this simplest version of the GBT can be 
expressed as \cite{carmo}
\begin{equation}
\int \int_T \mathcal{K} dS + \sum^{N}_{a\,=\,1} \int_{\partial T_a} \kappa_g dl + \sum^{N}_{a\,=\,1} \theta_a = 2\pi,
\label{eqn.15}
\end{equation}
where $T$ is a two-dimensional orientable surface shown in Fig.\ \ref{fig.2}. 
The boundaries of the surface $\partial T_a$ ($a = 1,2,...,N$) are 
differentiable curves. $\theta_a$ are the jump angles between the curves. 
$\mathcal{K}$ is the Gaussian curvature of the orientable surface $T$, 
$\kappa_g$ is the geodesic curvature of $\partial T_a$, $dS$ is the 
infinitesimal area element of the surface and $dl$ is the infinitesimal line 
element along the boundary. Also, $dl>0$ for prograde motion of photons and 
for retrograde motion $dl<0$. The sign of $dl$ is chosen to be consistent with 
the orientation of the surface $T$.

At this stage, we should note that the quadrilateral 
$\stackrel{\infty}{R}\!\square\!\stackrel{\infty}{S}$ 
shown in Fig.~\ref{fig.1} is embedded in a curved space $\mathcal{M}^{(3)}$ 
and consists of a spatial curve for a light ray from the source to the 
receiver, two outgoing radial lines from $R$ and from $S$, and a circular arc 
segment $C_r$ with the coordinate radius $r_C$ ($r_C \rightarrow \infty$). 
As clear from the figure, within the asymptotically flat spacetime, 
$\kappa_g \rightarrow 1/r_C$ and $dl \rightarrow r_C\, d\phi$ as 
$r_C \rightarrow \infty$ \cite{gibbons}. Hence, the deflection angle can be 
defined in the domain $\stackrel{\infty}{R}\!\square\!\stackrel{\infty}{S}$ as
\begin{equation}
\hat{\Theta} = \Psi_R - \Psi_S + \phi_{RS} = - \int \int_{\stackrel{\infty}{R}\square\stackrel{\infty}{S}} \mathcal{K}\, dS 
\label{eqn.16}
\end{equation}
The separation angle $\phi_{RS}$ for our system can be obtained by 
integrating Eq.~\eqref{eqn.14} as given by
\begin{equation}
\phi_{RS} = 2\int_{0}^{u_0}\frac{du}{\sqrt{F(u)}},
\label{eqn.16a}
\end{equation} 
where $u_0$ is the inverse of the distance of closest approach. As in the 
Ishihara $\textit{et al.}$ method, if we consider that the source and the 
receiver are at finite distances from each other, then deflection angle can 
be written as
\begin{equation}
\hat{\Theta} = \Psi_R - \Psi_S + \int_{u_R}^{u_0} \frac{du}{\sqrt{F(u)}} + 
\int_{u_S}^{u_0} \frac{du}{\sqrt{F(u)}}. 
\label{eqn.17}
\end{equation}
Again, for the metric \eqref{eqn.4} using Eq.~\eqref{eqn.12}, we obtain,
\begin{equation}
\begin{split}
\hspace{0.5cm}\Psi_R - \Psi_S &= \arcsin(\xi u_R) + \arcsin(\xi u_s) - \pi + \frac{\xi(\Lambda + c_0 M)}{2} 
\left[\frac{u_R^{-1}}{\sqrt{1 - \xi^2 u_R^2}} + \frac{u_S^{-1}}{\sqrt{1 - \xi^2 u_S^2}}\right] 
+ \xi M \left[\frac{u_R}{\sqrt{1 - \xi^2 u_R^2}}\right. \\
&\left.+\, \frac{u_S}{\sqrt{1 - \xi^2 u_S^2}} \right]\!
 + \frac{\xi M \Lambda}{2} \left[\frac{1 - 2\xi^2 u_R^2}{(1 - \xi^2 u_R^2)^{3/2}} + 
\frac{1 - 2\xi^2 u_S^2}{(1 - \xi^2 u_S^2)^{3/2}}\right] + \frac{\xi M c_0 \Lambda}{4} 
\left[\frac{u_R^{-3}}{(1 - \xi^2 u_R^2)^{3/2}} + \frac{u_S^{-3}}{(1 - \xi^2 u_S^2)^{3/2}}\right] \\
& - \frac{\xi^3 M c_0 \Lambda}{2}\left[\frac{u_R^{-1}}{(1 - \xi^2 u_R^2)^{3/2}} + \frac{u_S^{-1}}{(1 - \xi^2 u_S^2)^{3/2}}\right] + \mathcal{O}(M^2, M\Lambda^2,\Lambda^2).
\end{split}
\label{eqn.18}
\end{equation}
Here, it can be seen that the expansion of $\Psi_R - \Psi_S$ becomes divergent 
at $u_R \rightarrow 0$ and $u_S \rightarrow 0$ due to the fact that our 
spacetime is non-asymptotically flat. Hence, this series Eq.~(\ref{eqn.18}) 
must be used only within a certain limit of finite radius of convergence.
Moreover, for the metric (\ref{eqn.4}) the function $F(u)$ can be expressed as
\begin{equation}
F(u) = \frac{1}{\xi^2} - u^2 + 2 M u^3 + (\Lambda + c_0 M).
\label{eqn.19}
\end{equation}
Hence, the angle $\phi_{RS}$ for the metric (\ref{eqn.4}) is obtained as
\begin{equation}
\begin{split}
\phi_{RS} &= \pi - \arcsin(\xi u_R) - \arcsin(\xi u_S) + \frac{M}{\xi} 
\left[\frac{2 - \xi^2 u_R^2}{\sqrt{1 - \xi^2 u_R^2}} + \frac{2 - \xi^2 u_S^2}{\sqrt{1 - \xi^2 u_S^2}}\right]\\[2pt] 
&+ \frac{\xi^3 (\Lambda + c_0 M)}{2}\left[\frac{u_R}{\sqrt{1 - \xi^2 u_R^2}}
 + \frac{u_S}{\sqrt{1 - \xi^2 u_S^2}}\right] + \frac{\Lambda \xi M}{2}\left[\frac{2 - 3\xi^2 u_R^2}{(1 - \xi^2 u_R^2)^{3/2}} + 
\frac{2 - 3\xi^2 u_S^2}{(1 - \xi^2 u_S^2)^{3/2}}\right]\\[2pt]
&  - \frac{\Lambda \xi^5 c_0 M}{4} 
\left[\frac{3u_R - 2\xi^2 u_R^3}{(1 - \xi^2 u_R^2)^{3/2}} + \frac{3u_S - 2\xi^2 u_S^3}{(1 - \xi^2 u_S^2)^{3/2}}\right] + \mathcal{O}(M^2, M\Lambda^2, \Lambda^2).
\end{split}
\label{eqn.20}
\end{equation}
Using Eqs.~(\ref{eqn.18}) and (\ref{eqn.20}), we finally get the deflection 
angle of light for our considered black hole as
\begin{equation}
\begin{split}
\hat{\Theta} &= \frac{\xi (\Lambda + c_0 M)}{2} \left[\frac{1 + \xi^2 u_R^2}{u_R \sqrt{1 - \xi^2 u_R^2}} 
+ \frac{1 + \xi^2 u_S^2}{u_S \sqrt{1 - \xi^2 u_S^2}}\right] + \xi M
\left[\frac{u_R}{\sqrt{1 - \xi^2 u_R^2}} + \frac{u_S}{\sqrt{1 - \xi^2 u_S^2}}\right] \\
&\;\;\; + \frac{\xi M \Lambda}{2} \left[\frac{3 - 5\xi^2 u_R^2}{(1 - \xi^2 u_R^2)^{3/2}} + 
\frac{3 - 5\xi^2 u_S^2}{(1 - \xi^2 u_S^2)^{3/2}}\right] + \frac{\xi c_0 M \Lambda}{4} 
\left[\frac{u_R^{-3}}{(1 - \xi^2 u_R^2)^{3/2}} + \frac{u_S^{-3}}{(1 - \xi^2 u_S^2)^{3/2}}\right] \\
&\;\;\; - \frac{\xi^3 c_0 M \Lambda}{2} \left[\frac{u_R^{-1}}{(1 - \xi^2 u_R^2)^{3/2}} + 
\frac{u_S^{-1}}{(1 - \xi^2 u_S^2)^{3/2}}\right] + \frac{M}{\xi} 
\left[\frac{2 - \xi^2 u_R^2}{\sqrt{1 - \xi^2 u_R^2}} + \frac{2 - \xi^2 u_S^2}{\sqrt{1 - \xi^2 u_S^2}}\right] \\
&\;\;\; - \frac{\xi^5 c_0 M \Lambda}{4} \left[\frac{3u_R - 2\xi^2 u_R^3}{(1 - \xi^2 u_R^2)^{3/2}} 
+ \frac{3u_S - 2\xi^2 u_S^3}{(1 - \xi^2 u_S^2)^{3/2}} \right] + \mathcal{O} (M^2, M\Lambda^2, \Lambda^2).
\end{split}
\label{eqn.21}
\end{equation}
By virtue of Eq.~\eqref{eqn.18} few terms in the above expression may diverge 
in the far distance limit, $u_R \rightarrow 0$, $u_S \rightarrow 0$. As 
mentioned earlier, this is due to the reason that the spacetime we have 
considered here is non-asymptotically flat, similar to the Kottler spacetime 
\cite{kottler} used by Ishiahara {\textit{et al.}} \cite{ishihara}. As 
discussed in \cite{ishihara}, we can state that this divergence in the 
deflection angle in the far distance limit does not matter as the limit 
$u_R \rightarrow 0$, $u_S \rightarrow 0$ is not applicable for astronomical 
observations. Also, the effect of the boundary term contribution 
$\Lambda_{eff}=\Lambda+c_0 M$ on the deflection angle can be seen in 
Eq.~(\ref{eqn.21}). It is observed that the deflection angle will increase 
with an increase in the effective cosmological constant term. Thus, the 
boundary term coming from $f(\mathcal{T}, \mathcal{B})$ gravity has 
significant effect on the deflection angle. Further, from Eq.~(\ref{eqn.21}) 
for $\Lambda = c_0 = 0$, we can arrive at
\begin{equation}
\hat{\Theta} \simeq \frac{M}{\xi} \left[\frac{2 - \xi^2 u_R^2}{\sqrt{1 - \xi^2 u_R^2}} + 
\frac{2 - \xi^2 u_S^2}{\sqrt{1 - \xi^2 u_S^2}}\right],
\label{eqn.22}
\end{equation}
which at the far distance limit ($u_R \rightarrow 0$, $u_S \rightarrow 0$), 
reduces to the deflection angle in the Schwarzschild case,
\begin{equation}
\hat{\Theta} \simeq \frac{4M}{\xi}.
\label{eqn.23}
\end{equation}

\begin{figure}[h!]
\centerline{
\includegraphics[scale=0.35]{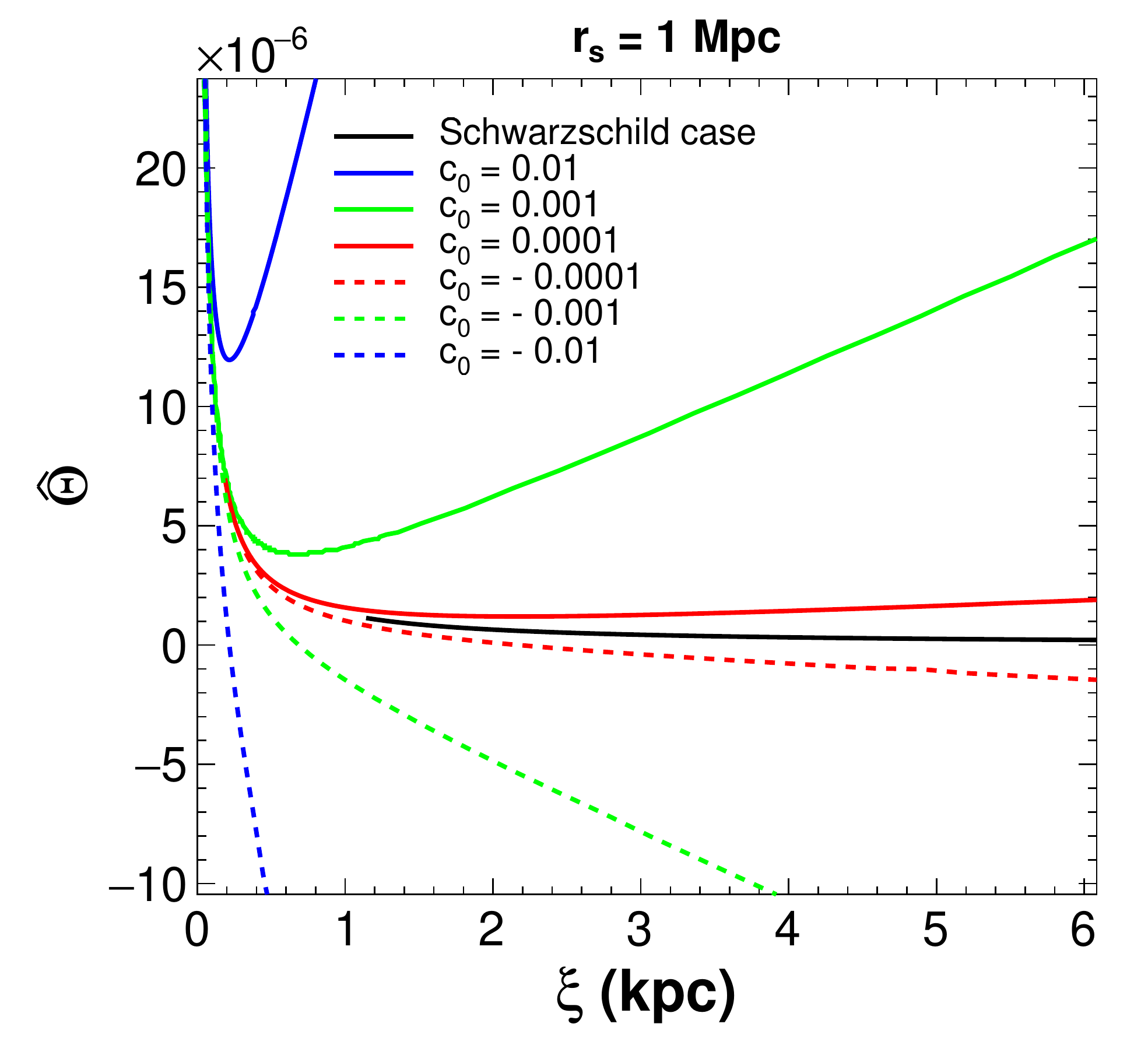}}\vspace{5mm}
\centerline{
\includegraphics[scale=0.35]{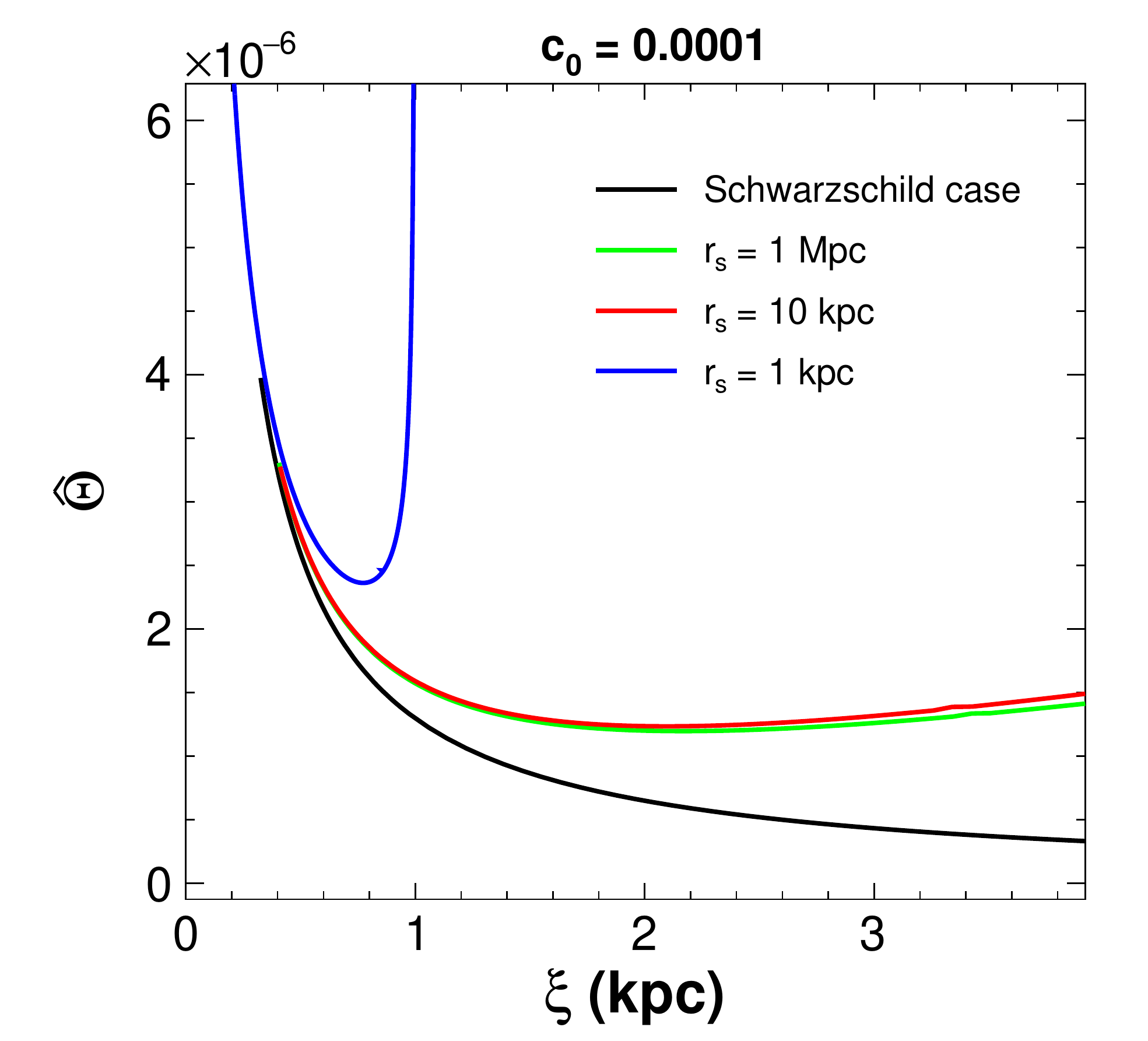}\hspace{5mm}
\includegraphics[scale=0.35]{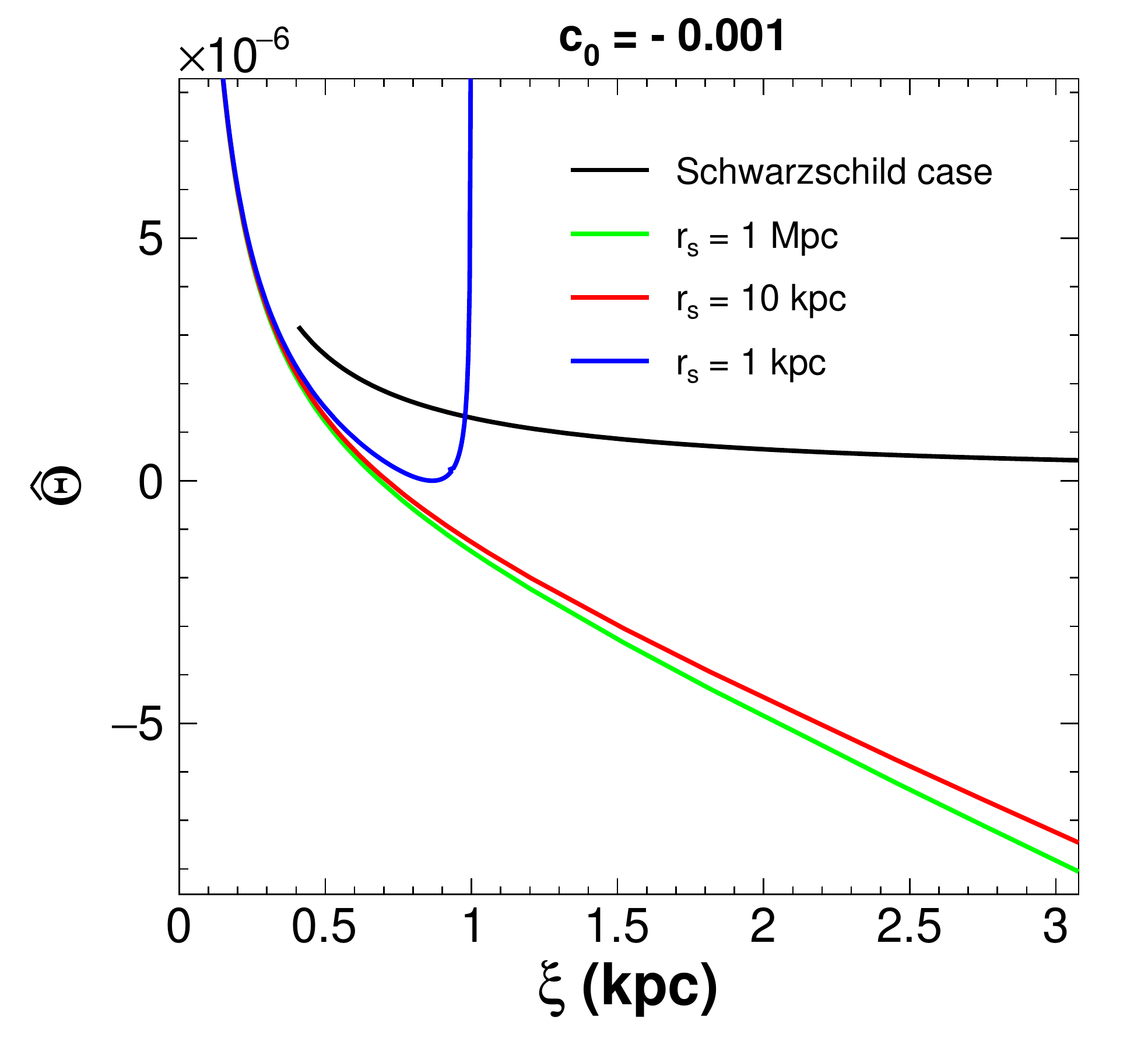}}
\caption{Deflection angle as a function of impact parameter for the case of
M$87^*$ black hole. In the first panel, we have considered $r_s = 1$ Mpc 
with three positive values of $c_0 = 0.01, 0.001, 0.0001$ in unit of 
kpc$^{-3}$ which are shown by the solid curves, and three negative values of 
$c_0 = -0.0001, -0.001, -0.01$ in the same unit which are represented by the 
dashed curves. In the second and the last panel, we have chosen $r_s = 1$ Mpc, 
$10$ kpc, $1$ kpc with $c_0 = 0.0001$ kpc$^{-3}$ and $c_0 = - 0.001$ 
kpc$^{-3}$ respectively.}
\label{fig.3}
\end{figure}

Now, as an example we will explore the deflection angle for M$87^*$, the 
central black hole candidate of M$87$ \cite{eht, akiyama} in the light of the 
above analysis. For this purpose we consider different values of the distance 
from the source $r_s$ and also different values of the model parameter $c_0$. 
The mass of this black hole is $\sim 6.5 \times 10^9 M_\odot \sim 10^{13}$ m. 
The distance of the receiver from M$87^*$ is its distance from the earth, 
which is $\sim 16$ Mpc $\sim 10^{23}$ m, and the recent cosmological data 
suggest the cosmological constant to be equal to $\Lambda = 10^{-52}$ m$^{-2}$ 
\cite{planck}. In Fig.~\ref{fig.3}, we plot the deflection angle as a function 
of the impact parameter, and compare our results with that of the 
Schwarzschild case. In the first panel of this figure, we have chosen a 
particular value of $r_s = 1$ Mpc. In this plot the solid curves depict the 
behaviour of the deflection angle for three positive values of the model 
parameter $c_0 = 0.01, 0.001, 0.0001$ and the dashed curves show the same for 
three negative values of the model parameter $c_0 = -0.0001, -0.001, -0.01$. 
All values of $c_0$ in this and rest of the calculations are used in the unit 
of kpc$^{-3}$. It can be 
seen that for $c_0 = 0.01$, the deflection angle decreases upto a certain 
value of the impact parameter and then abruptly increases. As the 
value of $c_0$ decreases, the deflection angle decreases with the impact 
parameter and increases slowly after a certain point. For $c_0 = 0.0001$, the 
curve almost overlaps with the Schwarzschild case upto a certain value of 
the impact parameter and then slowly increases. For the negative values of the 
model parameter, it is seen that the deflection angle decreases and becomes 
negative for a certain value of the impact parameter. For $c_0 = - 0.0001$, 
the curve almost overlaps with the Schwarzschild case and then slowly becomes 
negative. This negative deflection angle signifies the behaviour of 
light around the black hole considered in our study. It can be said from our results that 
at certain impact parameter values, for $c_0 < 0$ cases, the photons get repelled 
leading to negative deflection of light. This can be comprehended as the repulsive 
gravitational effect of the teleparallel gravity theory 
\cite{bahamonde_rev, hanafy}. This type of negative deflection angle has been 
found in various studies \cite{drgt, nakashi, kitamura, hagiwara, izumi, kar}. 
In the second and the last panel, we plot the deflection angle as a function 
of the impact parameter for $c_0 = 0.0001$ and $c_0 = - 0.001$ respectively, 
with three values of the distance from the source $r_s = 1$ Mpc, $10$ kpc, 
$1$ kpc. It is seen that if the source (say, a cluster of galaxies) 
is at a distance of $1$ kpc, the deflection angle decreases till a certain 
point and then abruptly diverges for both the cases of $c_0 = 0.0001$ and 
$c_0 = - 0.001$. However, for $c_0 = 0.0001$ the divergent behaviour is seen 
at a higher deflection angle than for $c_0 = - 0.001$. If we consider a galaxy 
cluster further away, say at $10$ kpc, there is a change in the way the 
deflection angle diverges for $c_0 = 0.0001$. Upto a certain value of the 
impact parameter, the deflection angle decreases in a way similar to the 
Schwarzschild case, but slowly becomes divergent after a particular value of 
the impact parameter. On the other hand, for $c_0 = - 0.001$, the deflection 
angle decreases and eventually becomes negative after a certain value of the 
impact parameter. For both the values of $c_0$, such similarities 
are also seen if we consider a cluster of galaxies at a distance of $1$ Mpc. 
Thus, the divergent behaviour of the deflection angle is observed at low 
impact parameter value when the galaxy cluster is nearer to the lens.

\section{Quasinormal modes}
\label{sec.4}
In this section, we will address the massless scalar perturbation in the 
spacetime of the black hole. We will assume that the test field exerts 
negligible influence on the black hole spacetime. To determine the quasinormal 
modes, we will derive Schr\"odinger-like wave equations taking into account 
the corresponding conservation relations of the concerned spacetime, which 
should be of Klein-Gordon type for the case of a scalar field. Two different 
methods, viz.\ the asymptotic iteration method (AIM) and the Pad\'e averaged 
6th order WKB approximation method, will be used to calculate the quasinormal 
modes. In this regard considering only the axial perturbations, we can express 
the perturbed metric as presented in \cite{lopez2020}:
\begin{equation} \label{pert_metric}
ds^2 = -\, |g_{tt}|\, dt^2 + r^2 \sin^2\!\theta\, (d\phi - p_1
dt - p_2 dr - p_3 d\theta)^2 + g_{rr}\, dr^2 +r^2 d\theta^2,
\end{equation}
where $p_1$, $p_2$ and $p_3$ define the perturbation introduced to the black 
hole spacetime and are functions of $t$, $r$ and $\theta$. The metric 
functions $g_{tt}$ and $g_{rr}$ represent the zeroth order terms and hence
are only functions of $r$.
\subsection{Scalar Perturbation}
We consider a massless scalar field near the previously established black 
hole. As it is considered that the effect of scalar field on the black hole 
spacetime is minimal, the perturbed metric Eq.\ \eqref{pert_metric} in this 
case can be expressed as 
\begin{equation}
ds^2 = -\,|g_{tt}|\, dt^2 + g_{rr}\, dr^2 +r^2 d \Omega^2.
\end{equation}
Now, for this case, it is feasible to write the Klein-Gordon equation in 
curved spacetime as
\begin{equation}  \label{scalar_KG}
\square \Phi = \dfrac{1}{\sqrt{-g}} \partial_\mu (\sqrt{-g} g^{\mu\nu}
\partial_\nu \Phi) = 0.
\end{equation}
With the help of this Eq.\ \eqref{scalar_KG} the quasinormal 
modes associated with the scalar perturbation can be described. For this 
purpose we decompose the scalar field $\Phi$ as follows:
\begin{equation}\label{scalar_field}
\Phi(t,r,\theta, \phi) = \dfrac{1}{r} \sum_{l,m} \psi_l(t,r) Y_{lm}(\theta,
\phi).
\end{equation}
In this equation, $Y_{lm}$ is the spherical harmonics with $l$ and $m$ are the
usual indices associated with it, $\psi_l(t,r)$ is the radial time-dependent 
wave function. Using Eqs.\ \eqref{scalar_KG} and \eqref{scalar_field} one can 
obtain the radial wave equation as
\begin{equation}  \label{radial_scalar}
\partial^2_{r_*} \psi_l(r_*) + \omega^2 \psi_l(r_*) = V(r) \psi_l(r_*).
\end{equation}
Here, in this expression $r_*$ is defined as 
\begin{equation}  \label{tortoise}
\dfrac{dr_*}{dr} = \sqrt{g_{rr}\, |g_{tt}^{-1}|}
\end{equation}
and is known as the tortoise coordinate. The term $V(r)$ represents the 
effective potential, whose explicit form is 
\begin{equation}  \label{Vs}
V(r) = |g_{tt}| \left( \dfrac{l(l+1)}{r^2} +\dfrac{1}{r \sqrt{|g_{tt}|
g_{rr}}} \dfrac{d}{dr}\sqrt{|g_{tt}| g_{rr}^{-1}} \right),
\end{equation}
here the term $l$ represents the multipole moment of the black hole's
quasinormal modes. In the present work, we will compute the quasinormal modes 
of the scalar perturbation of the black hole specified by the 
metric \eqref{eqn.4} using this potential expression.

\subsection{The asymptotic iteration method}

The AIM is an influential mathematical tool employed to solve differential 
equations numerically, especially to those that are intractable by analytical 
means. A critical area where AIM is particularly useful to apply is the 
quasinormal modes of black holes and other systems featuring a potential 
barrier \cite{AIM1, AIM2, AIM3, AIM4}. Quasinormal modes refer to the 
characteristic oscillations exhibited by a system after a disturbance and are 
instrumental in analyzing the stability and attributes of black holes. AIM 
utilizes a systematic iteration approach that facilitates the derivation of 
precise approximations to the quasinormal modes by converting the initial 
differential equation into a series of simpler equations that are readily 
solvable. The technique has proved successful in various physical systems and 
remains an active area of exploration.

%In the de Sitter or anti-de Sitter case, we define 
%$\mathcal{G} = 1/r$ following the Ref. \cite{AIM1}. 
With the previous definition $u = 1/r$ and following Ref.\ \cite{AIM1}, we 
obtain the master wave equation for our case as given by
\begin{equation}
\frac{d^2 \psi}{d u^2} + \frac{\mathcal{Z}'}{\mathcal{Z}} \frac{d\psi} {d u} + \left[ \frac{\omega ^2-\mathcal{Z} \left(-\frac{2 \left(c_0 M+\Lambda \right)}{u^2}+l (l+1)+2 M u \right)}{\mathcal{Z}^2}\right]\!\psi = 0, \label{masterxi}
\end{equation}
where the parameter $\mathcal{Z}$ is explicitly given by
\begin{equation}
\mathcal{Z}= -M \left(c_0+2 u^3\right)-\Lambda + u^2.
\end{equation}
Now, one needs to scale out the divergent characteristics of quasinormal modes 
at the cosmological horizon, which can be done by defining the wave function
$\psi(u)$ as
\begin{equation}
\psi(u) = e^{i\omega r_*}\, \mathcal{G} (u),  \label{SdScale}
\end{equation}
which gives us the privilege to have Eq.\ \eqref{masterxi} as
\begin{equation}
\mathcal{Z} \mathcal{G}'' + (\mathcal{Z}'- 2 i\omega)\mathcal{G}' - \left[ -\frac{2 \left(c_0 M+\Lambda \right)}{u^2}+l (l+1)+2 M u \right]\mathcal{G}=0. \label{youeq}
\end{equation}
Again, the correct quasinormal condition at the black hole horizon $u_1$ 
results,
\begin{equation}
\mathcal{G}(r_*) =  (u-u_1)^{-\frac{i\omega}{\kappa_1}}\chi(r_*),
\end{equation}
where $\kappa_1$ is given by
\begin{equation}
\kappa_1 = \left.\frac 1 2 \frac{d f}{ dr}\right|_{r\to r_1}\!\!\!\! =M u_1^2-\frac{c_0 M+\Lambda }{u_1}\,.
\end{equation}
In the above expressions, $u_1=1/r_1$, where $r_1$ is the 
event horizon radius of the black hole. With the new function $\chi(r_*)$, the
Eq.\ \eqref{youeq} takes the conventional format to be used in the iterations 
of AIM for the differential equation as 
\begin{eqnarray}
\chi''& = & \lambda_0(u)  \chi' + s_0(u) \chi\,,
\end{eqnarray}
where the parameters $\lambda_0$ and $s_0$ are defined as
\begin{align}
\lambda_0(u) &= -\frac{1}{\mathcal{Z}} \left[\mathcal{Z}'- \frac{2i\omega }{ \kappa_1(u-u_1)} - 2 i\omega\right], \\[5pt]
s_0 (u) &=  \frac{1}{\mathcal{Z}} \left[l(l+1)+ 2 M u -\frac{2 \left(c_0 M+\Lambda \right)}{u^2}\ +\frac{i \omega}{ \kappa_1(u-u_1)^2}\,\Big(\frac{i\omega}{\kappa_1} +1\Big) +(\mathcal{Z}'- 2 i\omega)\, \frac{i\omega}{ \kappa_1(u-u_1)}\right].
\end{align}
We shall use this differential equation and follow the Ref.\ \cite{AIM1} to 
calculate the scalar quasinormal modes for the case of our black hole.

\subsection{The Pad\'e averaged WKB approximation method}
Besides the AIM, in this study we use the Pad\'e averaged sixth order 
WKB approximation method to calculate the quasinormal modes of black hole
defined by the metric \eqref{eqn.4}, as mentioned earlier. In this sixth order 
WKB method, the expression of oscillation frequency $\omega$ of GWs can be 
given by
\begin{equation}
\omega = \sqrt{-\, i \left[ (n + 1/2) + \sum_{k=2}^6 \bar{\Lambda}_k \right] \sqrt{-2 V_0''} + V_0}\,,
\end{equation}
where $n = 0, 1, 2\hdots$, $V_0 = V(r)|_{r\, =\, r_{max}}$ and 
$V_0'' = \dfrac{d^2 V}{dr^2}|_{r\, =\, r_{max}}$. Here $r_{max}$ is the 
position at which the potential $V(r)$ has its maximum value. 
$\bar{\Lambda}_k$ are the correction terms and the explicit forms of these 
correction terms, as well as the Pad\'e averaging recipe, can be found in 
Ref.s \cite{Schutz,Will_wkb,Konoplya_wkb,Maty_wkb}.

\begin{table}[ht!]
\begin{center}
\caption{Quasinormal modes from the black hole defined by the metric 
\eqref{eqn.4} for different values of the multipole moment $l$ with the 
overtone number $n=0$ obtained by using the AIM (with $91$ iterations) and
the $6$th order Pad\'e averaged WKB approximation method. In this calculation 
we choose the model parameters $c_0 = -0.01$ and $\Lambda = 0.002$. Here we 
use $M=G=c=\hbar=1$ unit system.}
\vspace{0.3cm}
\begin{tabular}{cccccc}
\hline\\[-10pt]
%\multicolumn{1}{|l}{$l$} & \multicolumn{1}{l}{$M(\omega_R - i \omega_I)$} & \multicolumn{1}{l}{$\Delta_{rms}$} & \multicolumn{1}{l|}{$\Delta_6$}  \\
$l$ & AIM & Pad\'e averaged WKB & $\Delta_{rms}$ & $\Delta_6$ & $\Delta_{m}$\\[2pt]  \hline
\\[-10pt] 
 $1$ & $0.3271473  -0.1000942 i$ & $0.329269 -0.102651 i$ & $0.00743811$ & $0.000107719$ & $0.97115\%$ \\
 $2$ & $0.5377497  -0.1048303 i$  & $0.538571 -0.104694 i$ & $0.00234137$ & $0.0000262313$ & $0.15196 \%$ \\
 $3$ & $0.7481478  -0.1053878 i$ & $0.747713 -0.105344 i$ & $0.000294378$ & $5.808299\times10^{-6}$ & $0.05784 \%$ \\
 $4$ & $0.9592543  -0.1056549 i$ & $0.959321 -0.105657 i$ & $8.032384\times10^{-6}$ & $2.580773\times10^{-6}$ & $0.00691 \%$ \\
 $5$ & $1.1707337  -0.1057988 i$ & $1.171000 -0.105793 i$ & $0.000233917$ & $1.280549\times10^{-6}$ & $0.02266\%$ \\[2pt] \hline  
\end{tabular}
\label{Table01}
\end{center}
\end{table}

In Table \ref{Table01}, we list the quasinormal modes for different values of 
the multipole moment $l$ with overtone number $n=0$. In the second column, the 
quasinormal modes obtained from the AIM with $91$ iterations are listed and in 
the third column, the quasinormal modes obtained from the 6th-order Pad\'e 
averaged WKB approximation method are shown. In this table, $\Delta_{rms}$ 
represents the rms error associated with the Pad\'e averaged $6$th order WKB 
approximation method and $\Delta_6$ provides a measurement of the error from 
two nearby approximation orders defined as
\begin{equation}
\Delta_6 = \dfrac{|\omega_7 - \omega_5|}{2},
\end{equation}
where $\omega_5$ and $\omega_7$ respectively represent the quasinormal modes 
calculated by using the Pad\'e averaged $5$th order and $7$th order WKB 
approximation methods. In the last column, $\Delta_{m}$ represents the 
percentage deviation of quasinormal modes calculated by using WKB method from 
those calculated by using AIM. One can see that with an increase in the 
multipole moment $l$, the error associated with the quasinormal modes 
decreases. It is also seen that the AIM and the Pad\'e averaged 6th-order WKB 
approximation method provide very close results, that is the quasinormal modes 
from both methods are in good agreement with each other. Moreover, the 
agreement between these two methods becomes far better with the increasing
$l$ values. It is to be noted that as a characteristic of the WKB approximation 
method, it fails to provide significant results when the overtone number $n$ 
is greater than the multipole moment $l$ \cite{2021_gogoi,2022_gogoi, gogoi5}. 
Hence, for smaller values of $l$, it seems that AIM provides more accurate 
results than those obtained from the WKB method. One may further note that the 
quasinormal frequencies shown in Table \ref{Table01} are in geometric units 
with $M=1$. To convert them to physical units, we can use the following 
conversion formula \cite{Ferrari}:
\begin{equation}
f = \dfrac{32.26}{\eta}\, (M \omega_R) \; \text{kHz},
\end{equation} 
where $\eta=M/M_{\odot}$. As mentioned above, in the WKB approximation method, 
the errors decrease with an increase in $l-n$ and for $n>l$ the method fails 
to provide actual quasinormal frequencies with a reasonable accuracy 
\cite{konoplya_new}. Hence in the rest of the study, we shall consider $n=0$ 
and a higher value of $l$ for a better accuracy of the results of quasinormal 
frequency calculations.

\begin{figure}[h!]
\vspace{0.3cm}
\centerline{
   \includegraphics[scale = 0.5]{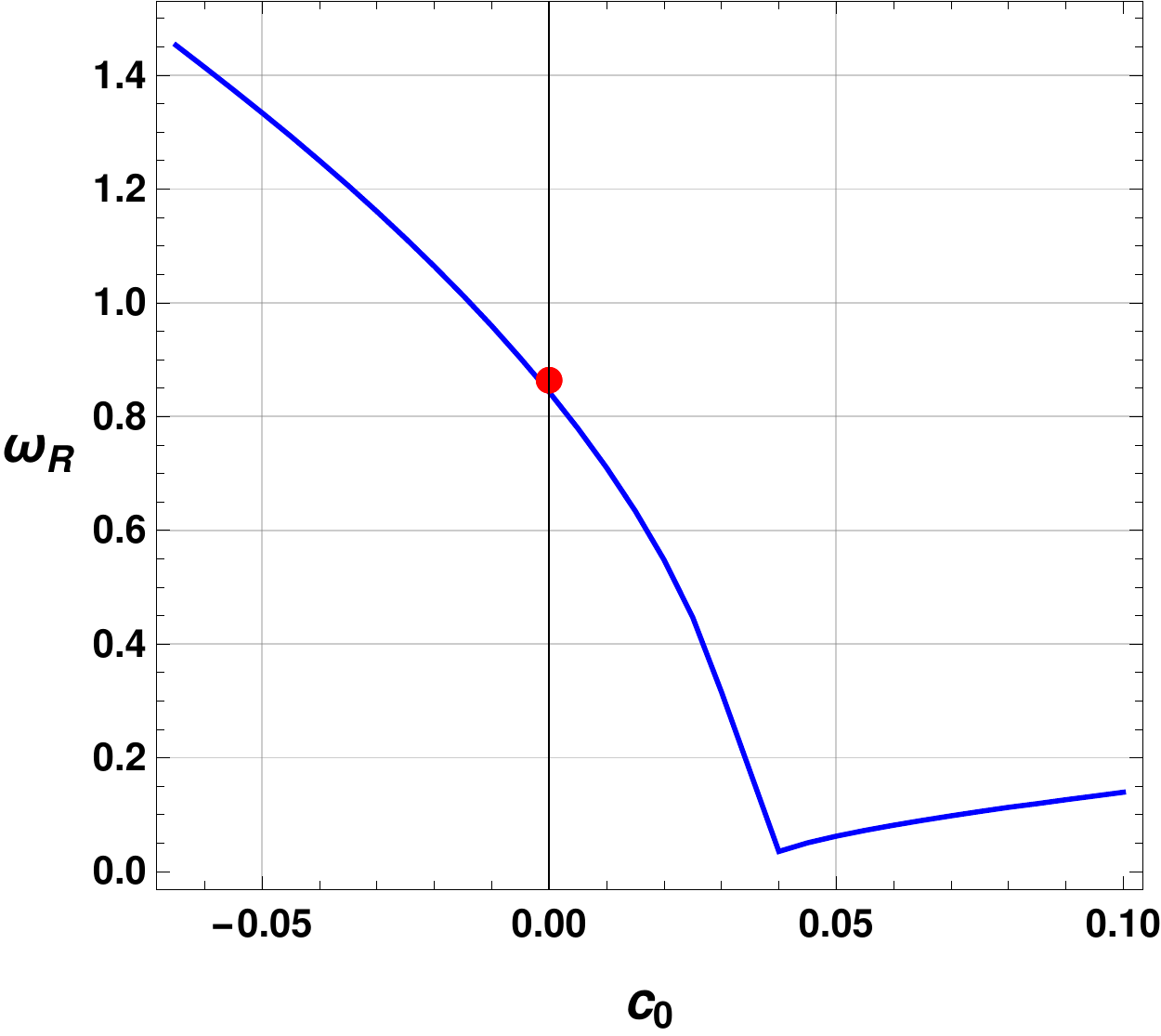}\hspace{0.5cm}
   \includegraphics[scale = 0.5]{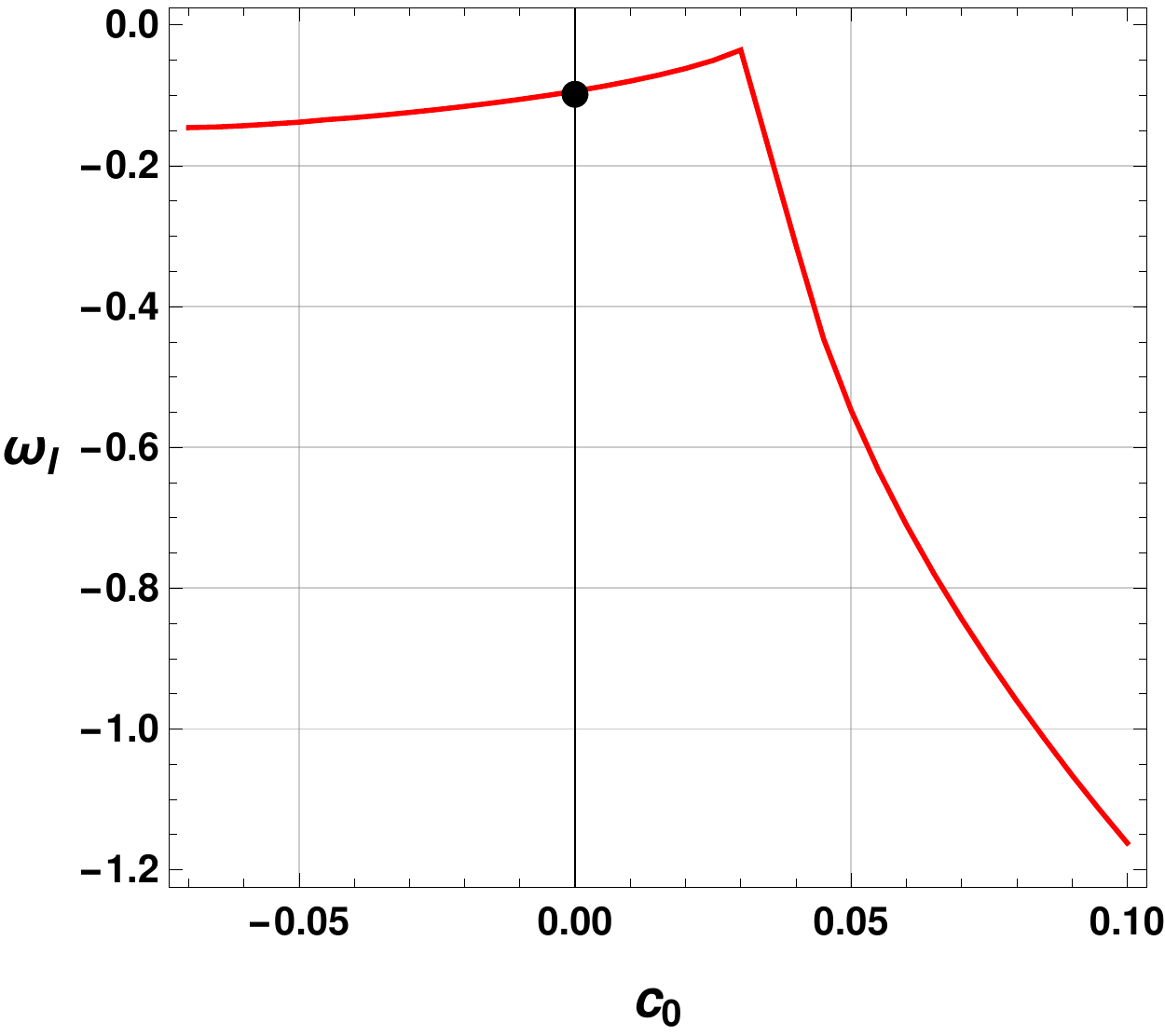}}
\vspace{-0.2cm}
\caption{Variation of real (on the left panel) and imaginary (on the right
panel) scalar quasinormal mode frequencies with the model parameter $c_0$
associated with the black hole metric \eqref{eqn.4} obtained by using
$n =0, l=4$ and $\Lambda = 0.002$. Solid circles in both plots denote quasinormal modes for asymptotically flat Schwarzschild black hole. Parameters are expressed in mass units
i.e. $M=1$.}
\label{Q01}
\end{figure}
In Fig.~\ref{Q01}, we plot the real quasinormal frequencies on the 
left panel and the imaginary quasinormal frequencies on the right panel with 
respect to the model parameter $c_0$. For both the plots, we use the mass of 
the black hole $M=1$, overtone number $n=0$, and multipole moment $l=4$. One 
can see that with an increase in the value of $c_0$, the quasinormal frequency 
decreases non-linearly and approaches zero towards $c_0 = 0.04$. However, 
beyond $c_0 = 0.04$, oscillation frequencies of quasinormal modes or ringdown 
GWs start to increase very slowly. So, it seems that the positive values of 
the model parameter $c_0$ permit ringdown GWs of very large wavelengths, which 
will be difficult to detect experimentally. Similarly, the decay rate or the 
damping rate of the quasinormal modes also decreases non-linearly with an 
increase in the model parameter $c_0$. Near $c_0=0.04$, the decay rate also 
becomes very close to zero. However, beyond this point, the decay rate 
increases drastically up to $c_0=0.1$, representing highly damped GWs.

\subsection{Time domain profiles}

In the previous two subsections, we numerically calculated the quasinormal
modes and studied their behaviour with respect to the model parameter $c_0$. 
In this subsection, we shall deal with the time domain profiles of the scalar 
perturbation of the black spacetime. To obtain the time evolution profiles, we 
shall implement the time domain integration formalism \cite{gundlach}.
For this purpose, we define the wavefunction and potential as 
$\psi(r_*,t) = \psi(i \Delta r_*, j \Delta t) = \psi_{i,j} $ and 
$V(r(r_*)) = V(r_*,t) = V_{i,j}$. With these definitions we can express
radial wave Eq.\ \eqref{radial_scalar} as
\begin{equation}
\dfrac{\psi_{i+1,j} - 2\psi_{i,j} + \psi_{i-1,j}}{\Delta r_*^2} - \dfrac{%
\psi_{i,j+1} - 2\psi_{i,j} + \psi_{i,j-1}}{\Delta t^2} - V_i\psi_{i,j} = 0.
\end{equation}
Now, we set the initial conditions 
$\psi(r_*,t) = \exp \left[ -\,\dfrac{(r_*-k_1)^2}{2\sigma^2} \right]$ and 
$\psi(r_*,t)\vert_{t<0} = 0$, where $k_1$ and $\sigma$ are the median and 
width of the initial wave-packet, and then calculate the time evolution of
the scalar field as
\begin{equation}
\psi_{i,j+1} = -\,\psi_{i, j-1} + \left( \dfrac{\Delta t}{\Delta r_*}
\right)^{\!2}\! (\psi_{i+1, j + \psi_{i-1, j}}) + \left( 2-2\left( \dfrac{\Delta t}{\Delta r_*} \right)^{\!2} - V_i \Delta t^2 \right) \psi_{i,j}.
\end{equation}
Using the above iteration scheme and choosing a fixed value of
$\frac{\Delta t}{\Delta r_*}$, one can easily obtain the profile of $\psi$ 
with respect to time $t$. However, one should keep 
$\frac{\Delta t}{\Delta r_*}< 1$ so that the Von Neumann stability condition 
is satisfied during the numerical procedure.

\begin{figure}[h!]
\vspace{0.2cm}
\centerline{
   \includegraphics[scale = 0.8]{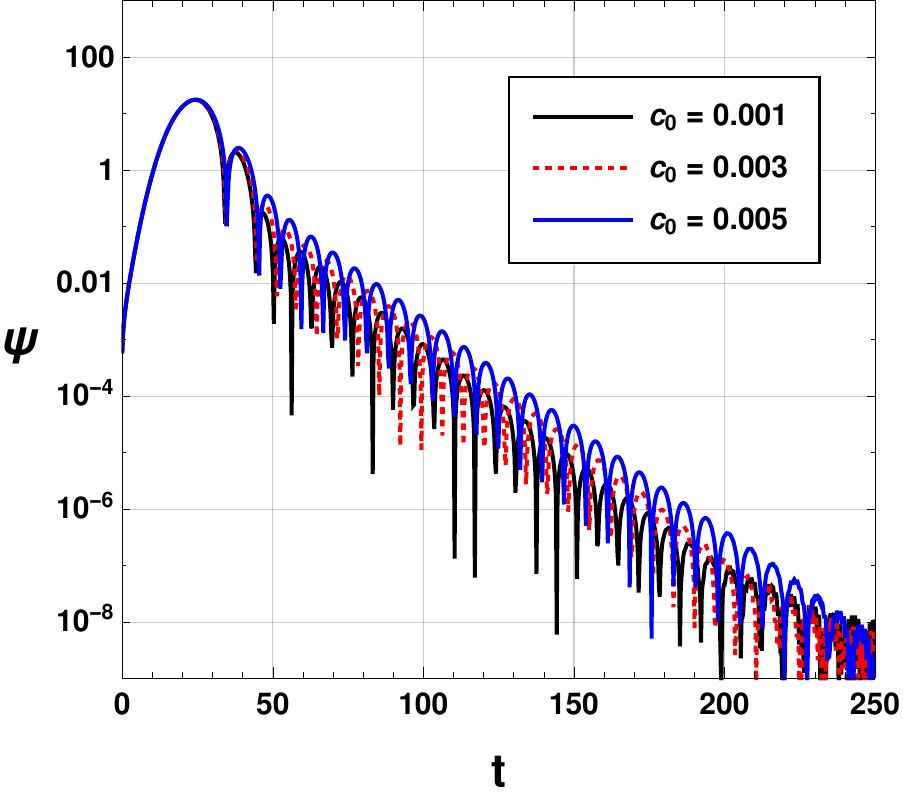}\hspace{0.5cm}
   \includegraphics[scale = 0.8]{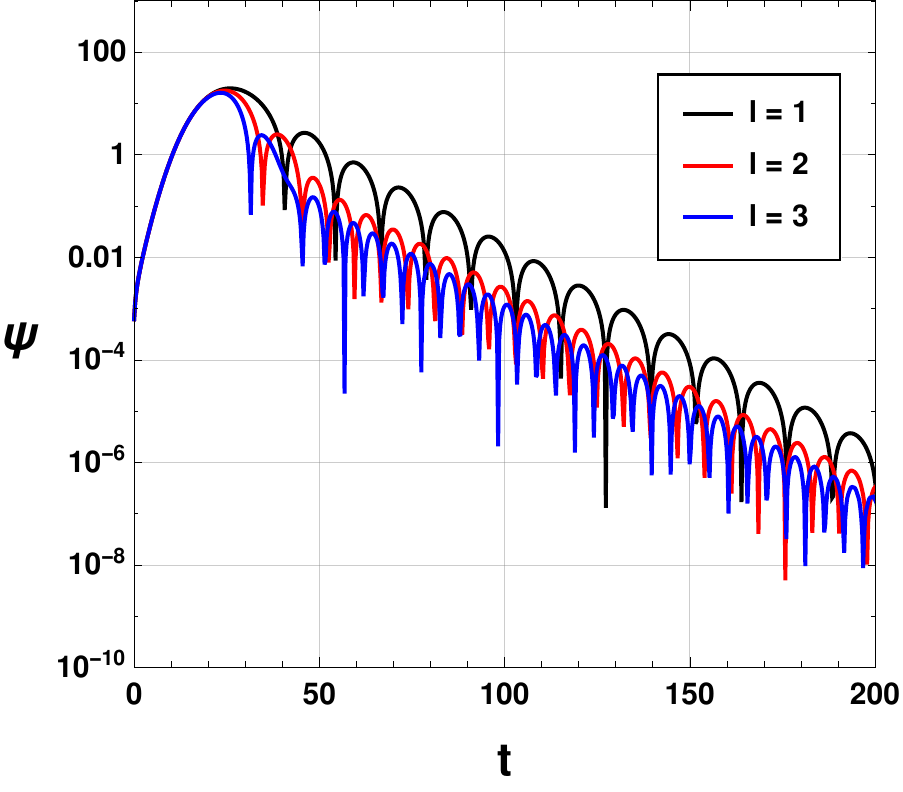}} \vspace{-0.2cm}
\caption{Time domain profiles with $n= 0$ and $\Lambda = 0.002$ for the 
massless scalar perturbation. On the left panel, we use $l=2$ and on the right 
panel, $c_0= 0.005$ is used. }
\label{time01}
\end{figure}

Fig.\ \ref{time01} shows the time evolution profiles of the massless scalar 
field perturbation in the black hole spacetime \eqref{eqn.4}. On the left 
panel, the time domain profiles are shown for different values of the model 
parameter $c_0$. One can see that with an increase in the value of $c_0$ from
$0.001$ to $0.005$, the oscillation frequencies decrease slowly. On
the right panel, we show the variation of the time domain profiles with 
different values of the multipole moment $l$. It is clear from the variation 
of the time domain profiles that with an increase in the value of the 
multipole moment $l$, both the oscillation frequency and the damping rate 
increase gradually. These results agree well with the numerically calculated 
quasinormal modes in the previous subsections.

\section{Optical behaviour of the black hole} \label{sec.5}
\subsection{Shadow}\label{sec.5A}
The shadow of a black hole is the dark area of spacetime that is surrounded by 
the event horizon of the black hole. Black holes are known to have an intense 
gravitational pull. The pull is so intense that once it passes the event 
horizon, even light cannot escape the gravitational attraction 
\cite{eslam, kumar}. As a result, the area of the black hole creates a 
shadowy dark zone on the background of the nearby matter or light. The 
dimension and shape of this shadow can provide vital insights into the 
characteristics of black holes and the nature of gravity. Recent astronomical 
and technical developments have made it possible to photograph black hole 
shadows, significantly advancing our understanding of these mysterious 
objects \cite{wei}.

\begin{figure}[h!]
\centerline{
   \includegraphics[scale = 0.55]{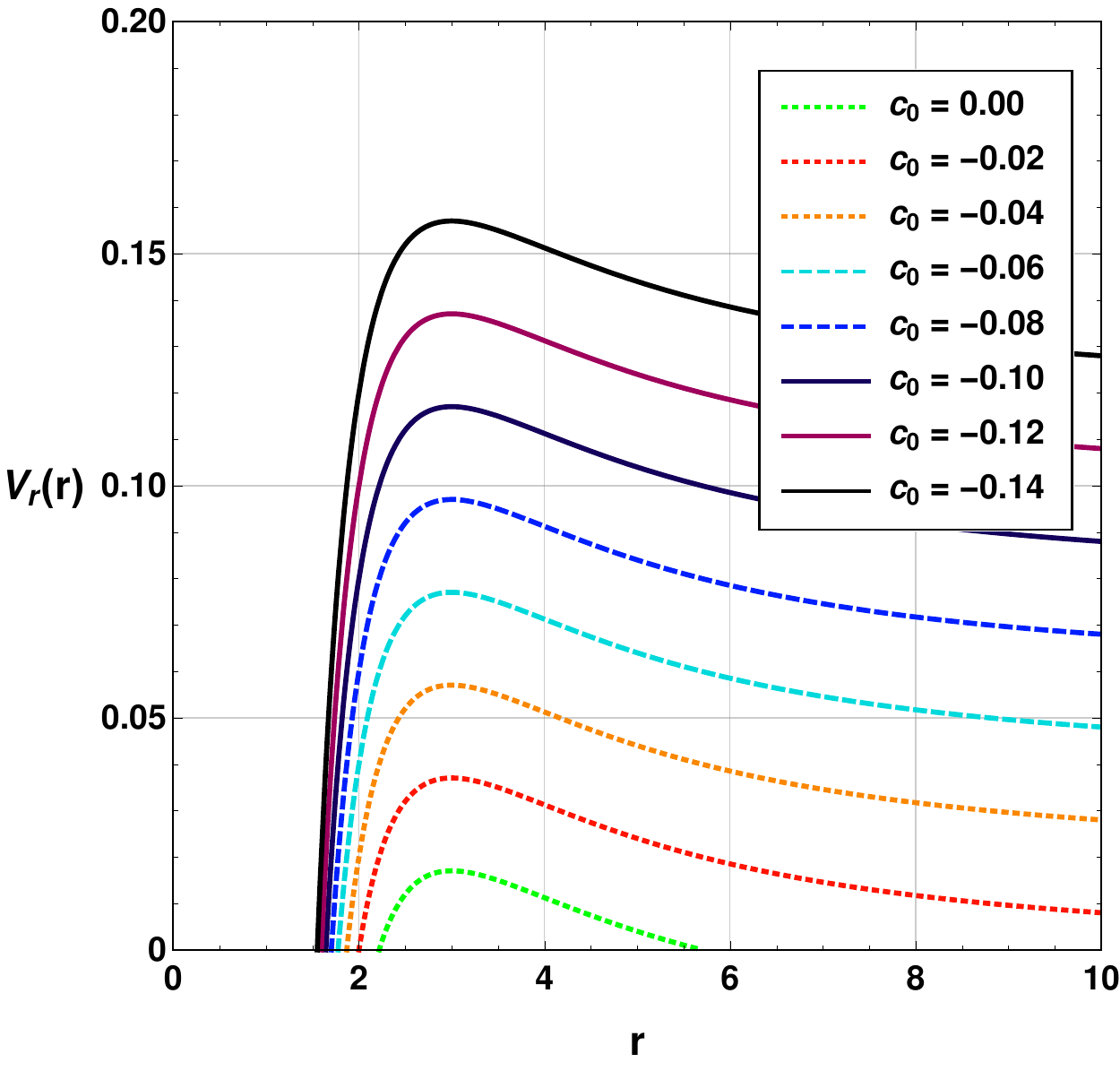}\hspace{0.5cm}
   \includegraphics[scale = 0.55]{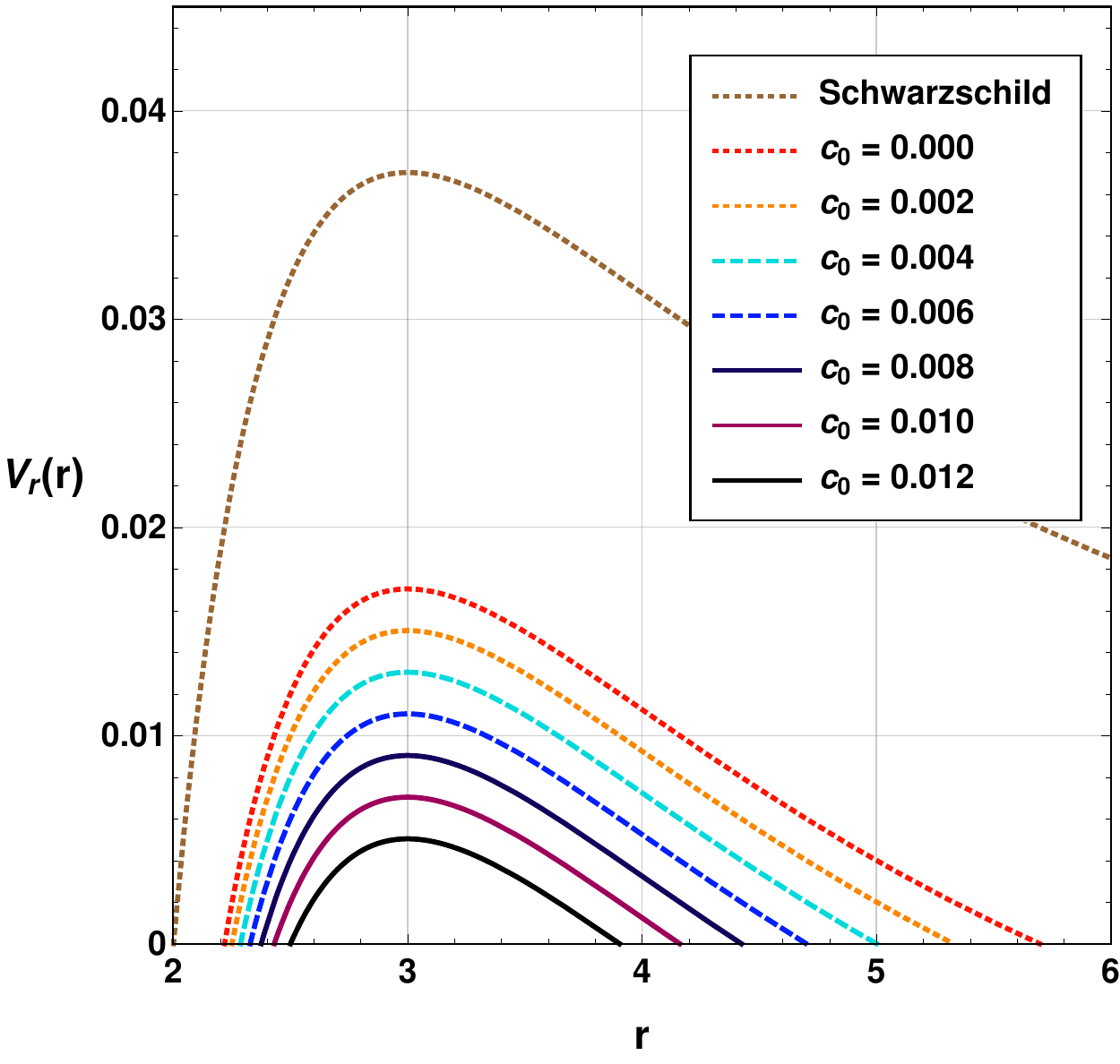}}
\vspace{-0.2cm}
\caption{Variation of the reduced potential $V_r(r)$ with respect to $r$. In 
these plots $M=1$ and $\Lambda = 0.02$ are used.}
\label{Shadow00}
\end{figure}

For the case of static and spherically symmetric spacetime metric, the 
Lagrangian of the form:
\begin{equation}
\mathcal{L}(x,\dot{x})=\frac{1}{2}\,g_{\mu\nu}\dot{x}^{\mu}\dot{x}^{\nu}, 
\end{equation} 
can be written as \cite{shnew02, ronit23}
\begin{equation}
\mathcal{L}(x,\dot{x})=\frac{1}{2}\left[-f(r)\,\dot{t}^{2}+\frac{1}{f(r)}\,\dot{r}^{2}+r^{2}\left(\dot{\theta}^{2}+\sin^{2}\theta\dot{\phi}^{2}\right)\right].
\end{equation}
Here, the derivative with respect to the proper time $\tau$ is indicated by 
the dot over the variables. The corresponding Euler-Lagrange equation is
\begin{equation}
\frac{d}{d\tau}\!\left(\frac{\partial\mathcal{L}}{\partial\dot{x}^{\mu}}\right)-\frac{\partial\mathcal{L}}{\partial x^{\mu}}=0.
\end{equation} 
For the present case of the study, choosing  the equatorial plane, i.e.\ 
$\theta=\pi/2$, the conserved energy $\mathcal{E}$ and angular momentum $L$ 
can be obtained using killing vectors $\partial/\partial \tau$ and 
$\partial/\partial \phi$ as \cite{ronit23}
\begin{equation}
\mathcal{E}=f(r)\,\dot{t},\quad L=r^{2}\dot{\phi}.
\end{equation}
The geodesic equation for the case of photon results in the relation,
\begin{equation}\label{eq22}
-f(r)\,\dot{t}^{2}+\frac{\dot{r}^{2}}{f(r)}\,+r^{2}\dot{\phi}^{2} = 0.
\end{equation}
In this Eq.\ \eqref{eq22}, using the conserved quantities {\it i.e.}, 
\ $\mathcal{E}$ and $L$ one can obtain the orbital equation of photon as given 
by \cite{shnew01}
%\begin{equation}\label{eff}
%\left(\frac{dr}{d\phi}\right)^{2}=r^{4} %\left[\frac{\mathcal{E}^{2}}{L^{2}}-\frac{f(r)}%{r^2}\right],
%\end{equation}
%Or,
\begin{equation}\label{eff}
\left(\frac{dr}{d\phi}\right)^{\!2}=V_{eff},
\end{equation}
where we define the right hand side of Eq.\ \eqref{eff} as an effective 
potential $V_{eff}$,  given by
\begin{equation}
V_{eff}= r^{4} \left[\frac{\mathcal{E}^{2}}{L^{2}}-\frac{f(r)}{r^{2}}\right].
\end{equation}
Furthermore, by writing Eq.\ \eqref{eff} in the form of a radial equation, 
one can have
\begin{equation}
    V_r(r) = \dfrac{1}{\xi^2} - \dot{r}^2/L^2,
\end{equation}
where the impact parameter $\xi$ is given as $\xi=L/\mathcal{E}$ and $V_r(r)$
is the reduced potential having the form:
\begin{equation}\label{pot}
V_r(r) = \frac{f(r)}{ r^2}.
\end{equation}
The study of this potential's behaviour with respect to the radial distance 
$r$ would be the most practical way to comprehend the nature of the photon 
sphere around the black hole spacetime we have taken into consideration. This 
potential governs the radial motion of photons in the black hole spacetime. 
In Fig.\ \ref{Shadow00}, we show the behaviour of this potential with respect
to $r$. One can see that the behaviour of the potential is not similar for 
positive and negative values of the model parameter $c_0$. For negative values 
of the parameter $c_0$, it is seen that the potential decreases slowly 
after reaching a peak point. For smaller values of the parameter $c_0$, the
potential increases to its peak value very swiftly and the peak of the 
potential increases gradually with a decreasing value of $c_0$. On the other 
hand, for positive values of the parameter $c_0$, the peaks are distinct,
reach to them comparatively slowly and the potential decreases 
drastically after reaching the peak with increasing values of $r$. In this 
case also, the peak value of the potential increases for lower values of the 
parameter $c_0$. Further, positive $c_0$ gives a lower peak value and negative
$c_0$ gives a higher peak value than that of the Schwarzschild case.  

Considering the turning point of the trajectory, given by $r=r_{ph}$, 
which is, in fact, the radius of the photon sphere or the light ring 
surrounding the black hole, one can determine the shadow of the black hole. 
At this turning point, the following conditions must be satisfied
\cite{18, synge, Luminet:1979nyg}:
\begin{equation} 
\left.\frac{dr}{d\phi}\right|_{r_{ph}}\!\!\!\!\!\!\!=0\;\; \text{or}\;\; \left.V_{eff}\right|_{r_{ph}}\!\!\!\!=0,\;\; \text{and}\;\;\; \left.\frac{d^2r}{d\phi^2}\right|_{r_{ph}}\!\!\!\!\!\!\!=0
\;\; \text{or}\;\; \left.V_{eff}^{\prime}\right|_{r_{ph}}\!\!\!\!=0.
\end{equation} 
Using the first condition, the impact parameter $\xi$ at the turning point can 
be obtained as
\begin{equation}
\frac{1}{\xi_{crit}^{2}}=\frac{f(r_{ph})}{r_{ph}^{2}}.
\label{impact}
\end{equation}
The radius of the photon sphere $r_{ph}$ can be determined using the second 
aforementioned condition and solving the equation: 
\begin{equation}
\left.\frac{d}{dr}\,\mathcal{A}(r)\right|_{r_{ph}}\!\!\!\!\!\!\! = 0.
\end{equation}
This equation can be explicitly written as
\begin{equation}
\frac{f^{\prime}(r_{ph})}{f(r_{ph})}-\frac{h^{\prime}(r_{ph})}{h(r_{ph})}=0,
\label{photon}
\end{equation}
where $\mathcal{A}(r)=h(r)/f(r)$ with $h(r)=r^{2}$. 
Thus from Eqs.\ \eqref{impact} and \eqref{photon} it is clear that the 
critical impact parameter is $\xi_{crit}=3 \sqrt{3} M/\sqrt{-27 c_0 M^3-27 \Lambda  M^2+1}$ and the photon sphere is located at $r_{ph}=3M$.
  
Now, to obtain the expression for shadow of the black hole, we rewrite 
Eq.\ \eqref{eff} with Eq.\ \eqref{impact} in terms of the function 
$\mathcal{A}(r)$ as 
\begin{equation}
\left(\frac{dr}{d\phi}\right)^{\!2}= h(r)f(r)\left(\frac{\mathcal{A}(r)}{\mathcal{A}(r_{ph})}-1\right).
\label{eq33}
\end{equation}
Using this Eq.\ \eqref{eq33}, the shadow radius can be determined. To this end, if one considers the angle between the light rays from a static observer at 
$r_0$ and the radial direction of the photon sphere as $\alpha$, then this 
angle can be calculated as \cite{Perlick:2021aok, 18}
\begin{equation}
\cot\alpha=\frac{1}{\sqrt{f(r)h(r)}}\left.\frac{dr}{d\phi}\right|_{r\,=\,r_{0}}\!\!\!\!\!\!\!\!\!\!\!.
\end{equation}
Together with Eq.\ \eqref{eq33}, above equation can be written as 
\begin{equation}
\cot^{2}\!\alpha=\frac{\mathcal{A}(r_{0})}{\mathcal{A}(r_{ph})}-1.
\end{equation}
Again, above equation can be rewritten using the relation $\sin^{2}\!\alpha=1/(1+\cot^{2}\!\alpha)$ as
\begin{equation}
\sin^{2}\!\alpha=\frac{\mathcal{A}(r_{ph})}{\mathcal{A}(r_{0})}.
\end{equation}
Substitution of the actual form of $\mathcal{A}(r_{ph})$ from 
Eq.\ \eqref{impact} and $\mathcal{A}(r_{0}) = r_0^2/f(r_0)$ the black hole's 
shadow radius for a static observer at $r_{0}$ is estimated as \cite{15s} 
\begin{equation} \label{shadow_exp}
R_{s}=r_{0}\sin\alpha=\sqrt{\frac{r_{ph}^2f(r_{0})}{f\left(r_{ph}\right)}}.
\end{equation}
In case of an asymptotically flat black hole and for a static observer at 
large distance, i.e.\ at $r_0 \rightarrow \infty$, $f(r_0) \rightarrow 1$. So 
for such an observer the shadow radius $R_s$ of this type of black hole 
becomes, 
\begin{equation}
R_{s} = \frac{r_{ph}}{\sqrt{f(r_{ph})}}.
\end{equation} 

The apparent form of the shadow of a black hole can be determined via the 
stereographic projection of the shadow from the black hole's plane to the 
observer's image plane with coordinates $(X,Y)$. These coordinates are defined 
as \cite{shnew03, ronit23} 
\begin{align}
 X & =\lim_{r_{0}\rightarrow\infty}\left(-\,r_{0}^{2}\sin\theta_{0}\left.\frac{d\phi}{dr}\right|_{r_{0}}\right),\\[5pt]
Y & =\lim_{r_{0}\rightarrow\infty}\left(r_{0}^{2}\left.\frac{d\theta}{dr}\right|_{(r_{0},\theta_{0})}\right),
\end{align}
where $\theta_{0}$ represents the angular position of the observer with respect
to the plane of the black hole.  
Depending on the value of $c_0$, the black hole spacetime can be de-Sitter (dS) or Anti de-Sitter (AdS) for a fixed value of $\Lambda$. It is to be noted that
dS solution gives rise to two horizons viz., event horizon and cosmological 
horizon, whereas AdS has only event horizon similar to the case of the
Schwarzschild black hole. In general, one may note that physical observers can 
be present between the event horizon and cosmological horizon with different 
physically allowed observer distance i.e.\ $r_1 < r_0 < r_\Lambda$, where 
$r_1$ and $r_\Lambda$ are the event and cosmological horizons respectively. 
Hence a variation in $r_0$ can have impacts on the appearance of the black 
hole shadow \cite{Perlick:2021aok}. On the other hand, for the asymptotically 
flat black holes with a static observer at large distance scenario as 
mentioned above, the apparent form of the shadow remains independent of the 
observer position. Therefore, in this investigation, we use the shadow 
radius expression \eqref{shadow_exp} to obtain the stereographic projections 
of the black hole shadow at some finite physical observer distance $r_0$.

\begin{figure}[h!]
\vspace{0.3cm}
\centerline{
   \includegraphics[scale = 0.5]{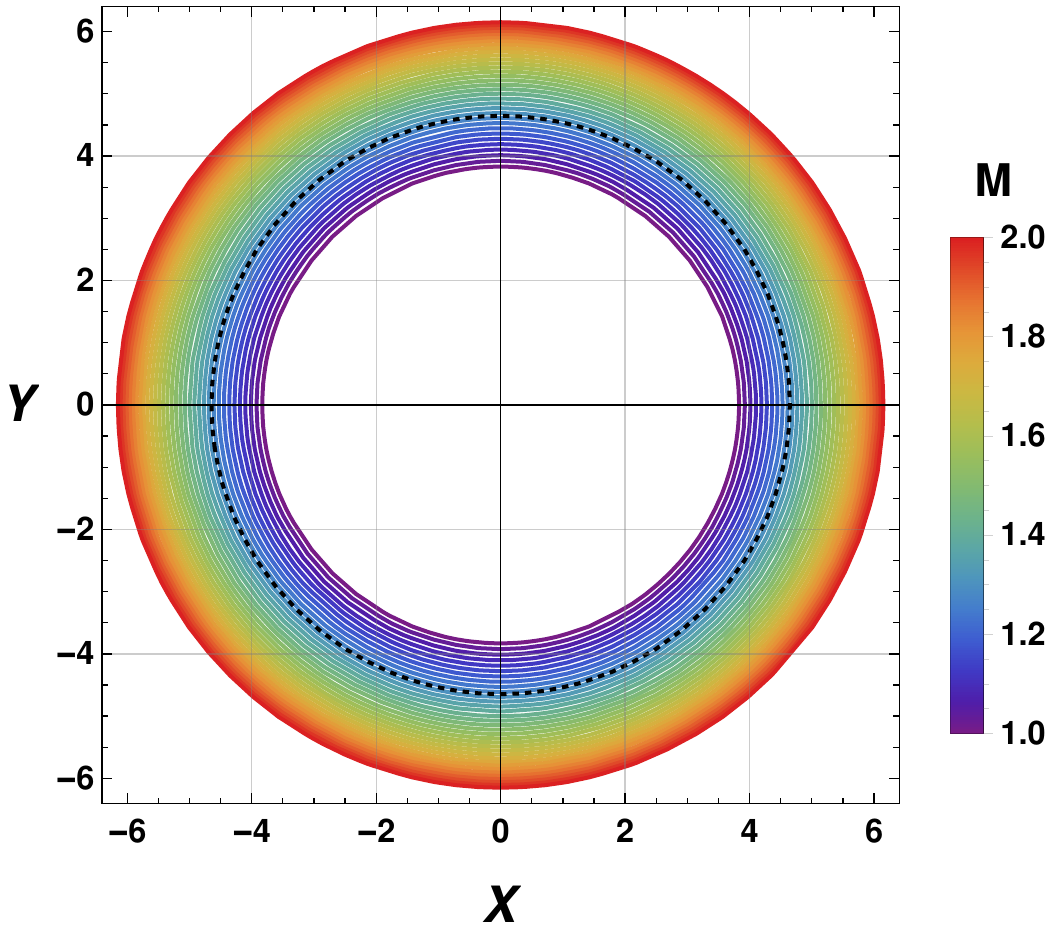}\hspace{1.0cm}
   \includegraphics[scale = 0.5]{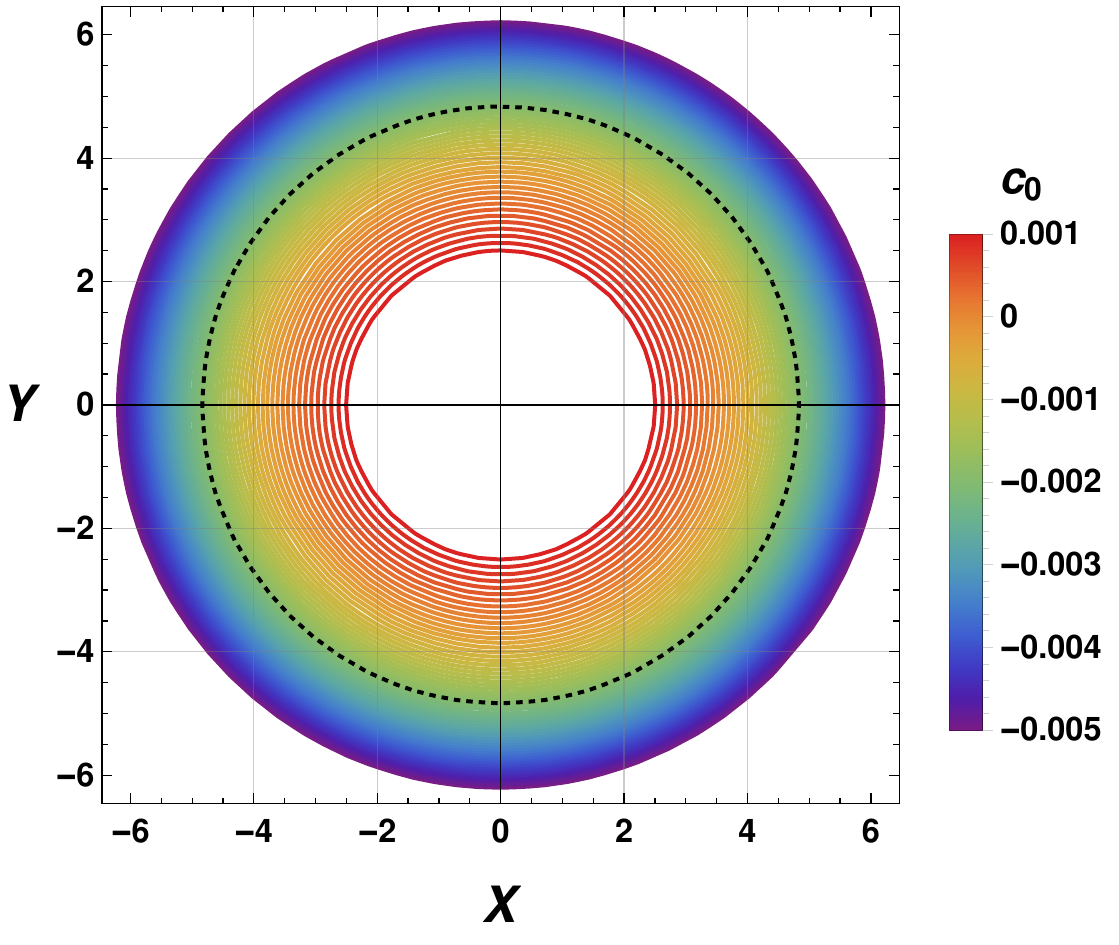}}
\vspace{-0.2cm}
\caption{Stereographic projections of the shadow of the black hole with 
$\Lambda = 0.002$. On the left panel, we use $c_0 = 0.001$ and $r_0 = 10$ and 
on the right panel, $M=1$ and $r_0 = 15$. The black dotted circle in both cases denote Schwarzschild black hole shadow radius at respective finite observer 
distances.}
\label{Shadow01}
\end{figure}

\begin{figure}[h!]
\vspace{0.3cm}
\centerline{
   \includegraphics[scale = 0.5]{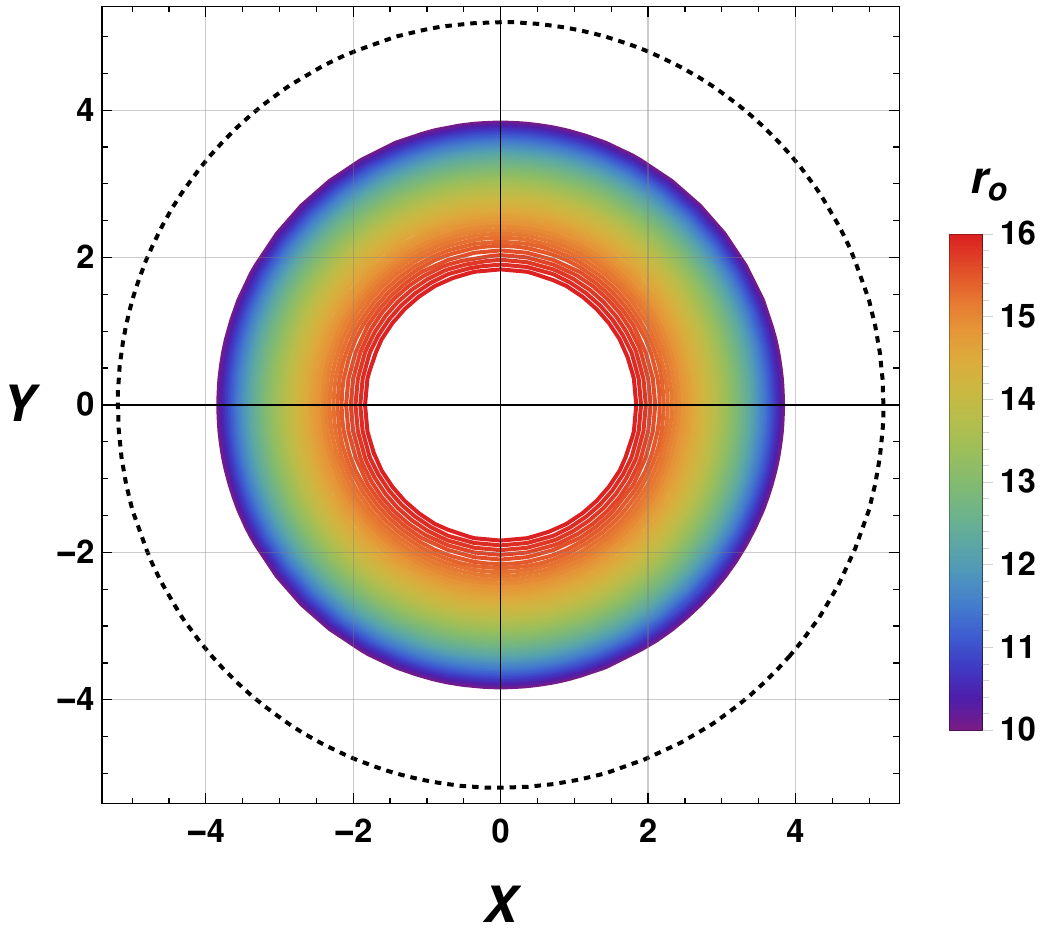}
   \hspace{1.0cm}
   \includegraphics[scale = 0.5]{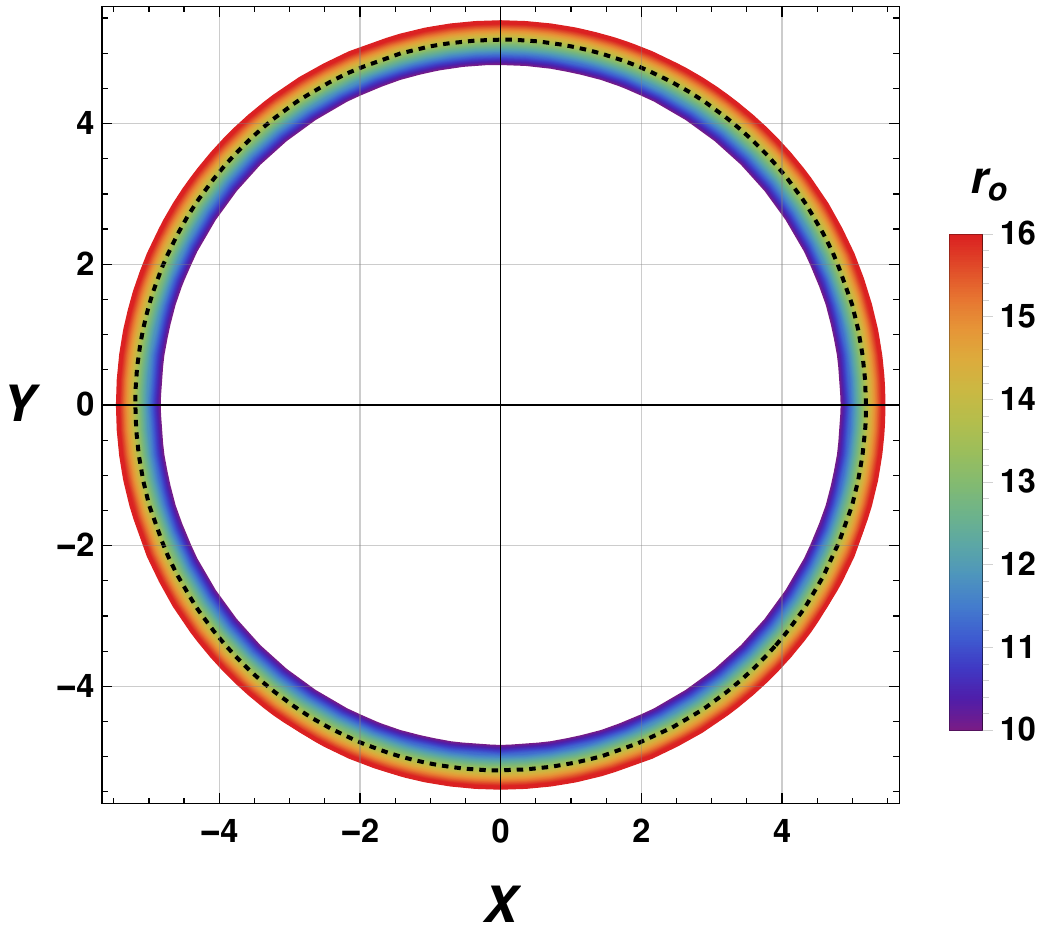}
}
\vspace{-0.2cm}
\caption{Stereographic projections of the shadow of the black hole with
$\Lambda = 0.002$ and $M = 1$. On the left panel, we use $c_0 = 0.001$ 
(dS case) and on the right panel, $c_0 = -0.003$ (AdS case). The black dotted 
circle in each plot denotes the Schwarzschild black hole shadow radius at 
infinite observer distance (i.e. at $r_0 \rightarrow \infty$).}
\label{Shadow02}
\end{figure}
 
For the black hole considered in this study, the stereographic projections of 
the black hole shadow at finite observer distances are shown in Fig.\ 
\ref{Shadow01}. On the left panel, we show the behaviour of the black hole 
shadow radius for different values of the black hole mass $M$. In this case, 
we observe that the shadow radius increases with an increase in the value of 
black hole mass $M$ at finite observer distance $r_0 = 10$. On the right panel, we show the dependency of the black hole shadow with the model parameter 
$c_0$. With an increase in the value of the parameter $c_0$, we observe a 
decrease in the size of the black hole shadow at finite observer distance 
$r_0 = 15$. That is, the size of the black hole shadow is bigger for the AdS 
ones than that of the dS cases. Finally, in Fig.\ \ref{Shadow02}, we show the 
dependency of the shadow radius with the observer distance $r_0$ for both
dS and AdS black holes considering $\Lambda = 0.002$ and $M = 1$. For the dS 
case (left panel) we choose the model parameter $c_0 = 0.001$ for which the 
black hole can have two horizons at $r = 2.02491$ (event horizon) and 
$r = 17.1606$ (cosmological horizon). The physical observer distance should 
lie within this range. We consider $r_0 = 10$ to $16$ to obtain the 
stereographic projections of the black hole shadow. For comparison, we have 
also plotted the Schwarzschild black hole shadow at infinite observer 
distance. It is clear from the figure that the observer distance plays a 
crucial role in depicting the black hole shadow radius. In this dS case, we 
see that with an increase in the observer distance $r_0$, the shadow radius 
decreases and they are smaller than the Schwarzschild black hole shadow. In the
case of AdS black holes we take the parameter $c_0 = - 0.003$ (right panel). It 
is seen that with an increase in the distance $r_0$ of the observer the 
shadow size increases and is comparable to that of the Schwarzschild black 
one, in contrast to the case of dS black holes.    

\subsection{Emission rate} \label{sec.5B}
By examining the black hole shadow, it is possible to investigate the emission 
of particles in the vicinity of the black hole. It has been demonstrated that 
for an observer located far away, the black hole shadow is indicative of its 
high-energy absorption cross-section \cite{Wei2013}. Generally speaking, in 
the case of a spherically symmetric black hole, the absorption cross-section 
exhibits oscillatory behaviour around a constant limiting value 
$\sigma_{lim}$ at extremely high energies. This limiting value $\sigma_{lim}$
corresponds to the geometrical cross-section of the photon sphere around the
black hole. Since the shadow provides a means of visually detecting a black 
hole, it is roughly equivalent to the area of the photon sphere 
($\sigma_{lim} \approx \pi R_{s}^{2}$). Thus the energy emission rate can be 
calculated by using the equation \cite{Wei2013},

\begin{equation}\label{Eqemission}
\frac{d^{2}E(\omega_p )}{dt\,d\omega_p }=\frac{2\pi^{3}\omega_p^{3}R_{s}^{2}}{e^{\omega_p/T}-1}, 
\end{equation}%
in which $\omega_p $ is the emission frequency, and $T$ is Hawking temperature 
given by
\begin{equation}
T= \dfrac{\hbar f'(r_h)}{4 \pi},  \label{EqTH}
\end{equation}
where $r_h$ is the horizon radius of the black hole.
\begin{figure}[h!]
\vspace{0.3cm}
\centerline{
   \includegraphics[scale = 0.7]{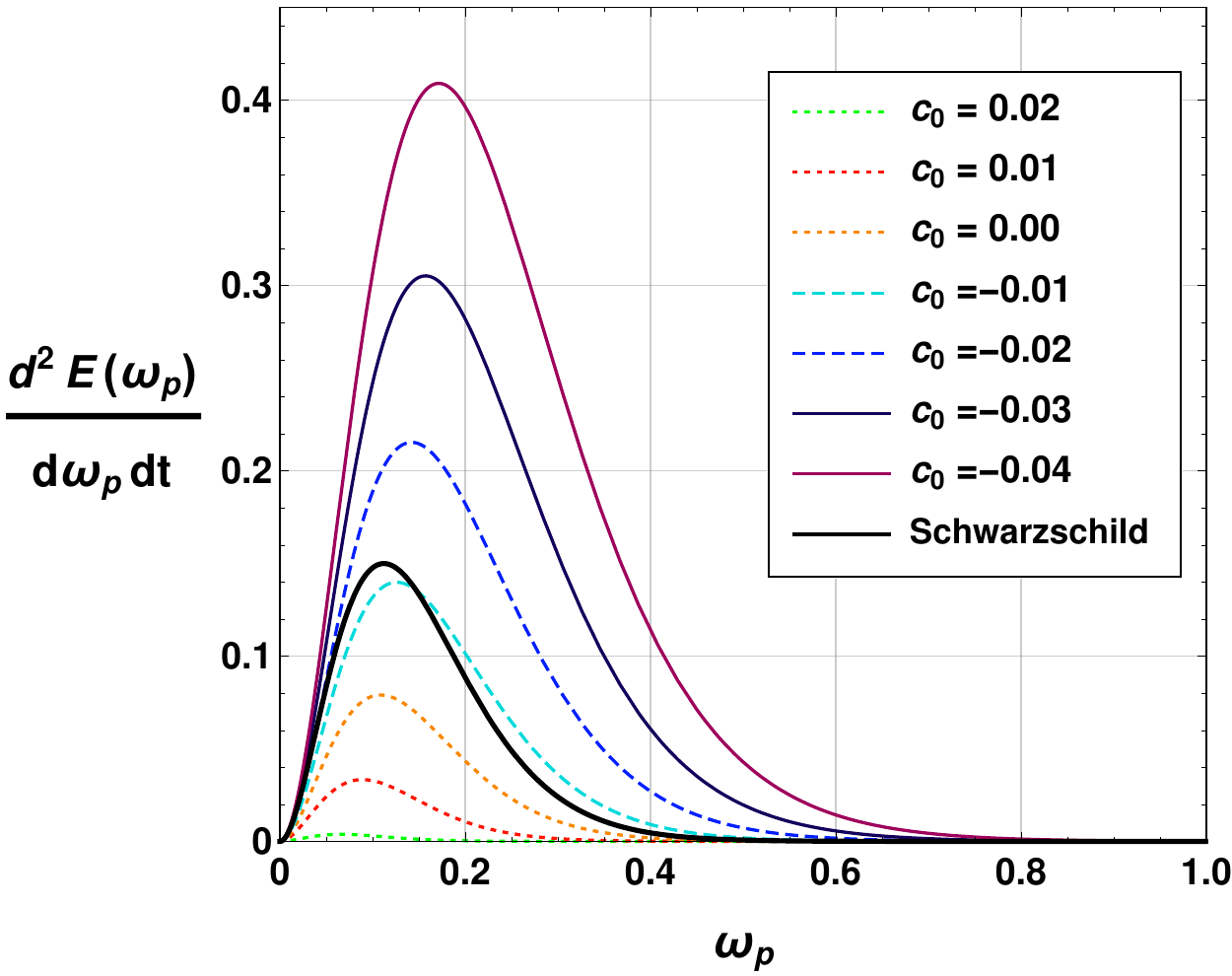}}
\vspace{-0.2cm}
\caption{Emission rate of the black hole with respect to emission frequency 
$\omega_p$ for $M=1$, $\Lambda = 0.002$ and observer distance $r_0 = 5$.}
\label{Emission}
\end{figure}
 
Using the expression of Hawking temperature \eqref{EqTH} in 
Eq.\ \eqref{Eqemission}, we show the variation of the emission rate with 
respect to $\omega_p$ for different values of $c_0$ in Fig.\ \ref{Emission}. 
One can see that with an increase in the value of the parameter $c_0$, the 
emission rate of the black hole decreases. Hence for smaller values of the 
parameter $c_0$, the black hole will evaporate more rapidly. Moreover, it 
is to be noted that for $c_0<-0.01$ the emission rate is greater than that of 
the Schwarzschild black hole, whereas it is less than Schwarzschild one for 
the positive values of $c_0$. For the $c_0 = -0.01$ case, the rate is very 
close to the Schwarzschild black hole. Here we consider the observer distance 
$r_0 = 5$, the cosmological constant $\Lambda = 0.002$ and the black hole mass 
$M = 1$. Thus in this figure the cases for $c_0\le -0.01$ correspond to the AdS 
black holes situations, whereas other values of $c_0$ correspond to the dS 
black holes.  
%which lies between the event horizon 
%and cosmological horizon for the black hole with dS case. It is 
%necessary because the emission rate of the black hole is shadow dependent 
%and from the previous investigation, we found that for dS case there are two 
%black hole horizons viz. event horizon and cosmological horizon. 

\section{Conclusion}
\label{sec.6}

In this work, we study a recently introduced static black hole solution in an 
extension of modified teleparallel gravity i.e., $f(\mathcal{T},\mathcal{B})$ 
modified gravity, which includes the function of the torsion scalar 
$\mathcal{T}$ and a related boundary term $B$. In this modified gravity theory, 
we compute the deflection angle of light by a non-asymptotically flat black 
hole. Then we study the quasinormal modes associated with the axial scalar 
perturbation in the background of the black hole.

Gibbons and Werner first introduced an alternative way to calculate the 
gravitational bending angle using the Gauss-Bonnet theorem. They first 
evaluated the deflection angle for a Schwarzschild black hole. Since then 
their work has been extended in various ways for different kinds of black 
holes. Few researchers have computed the deflection angle from the stationary 
black holes. Few have again considered finite distances between source and 
receiver, and derived the deflection angle by the static as well as stationary 
black holes. Some recent studies also considered black holes in modified 
gravity theories to obtain the deflection angle of light. In this work, we 
implemented the Ishihara \textit{et al.}~method to evaluate the deflection 
angle of light from the receiver point of view. This method does not depend 
on the asymptotic flatness. Hence, we have applied this method in the 
non-asymptotically flat black hole in $f(\mathcal{T},\mathcal{B})$ gravity. 
We have computed the bending angle considering the source and the receiver at 
a finite distance. It is found that when the source and the receiver are 
considered near to the lens object, the deflection angle becomes divergent. 
%However, this divergence becomes an issue only when the source or the receiver 
%is detected at the horizon. 
The boundary term coming from $f(\mathcal{T},\mathcal{B})$ gravity has 
a significant effect on the deflection angle. In the near future, we wish 
to extend our work to obtain the deflection angle of massive particles using 
Ishihara $\textit {et al.}$~method in the black hole as well as in wormhole 
backgrounds in different MTGs.

Quasinormal modes are some complex numbers related to the emission of GWs 
from the compact objects in the universe. The real part of these modes is 
related to the emission frequency and the imaginary part is related to the 
damping. In this work, we try to find a Schr{\"o}dinger type equation 
containing an effective potential, which allows us to compute the quasinormal 
modes associated with the massless scalar perturbations in the background 
of the considered black hole in the $f(\mathcal{T},\mathcal{B})$ gravity. For 
this purpose, we use the AIM and Pad\'e averaged sixth-order WKB approximation 
method. We obtained the quasinormal modes for different values of the 
multipole moment $l$ with the overtone number $n = 0$. It has been observed 
that for smaller values of $l$, the error associated with the quasinormal 
frequencies calculated with WKB method is higher in magnitude. With an 
increase in the value of $l$, the errors associated with the frequencies 
decrease significantly. The AIM stands in agreement with the results obtained 
by using the WKB method, confirming the validity of the results. Our study 
shows that the parameter $c_0$ has a significant impact on the quasinormal 
modes of the black hole. Moreover, the time domain profile analysis of 
evolution of massless scalar field perturbation in the black hole spacetime 
also confirms all these outcomes of quasinormal modes' calculations.  
However, for the observational constraints on the model from the quasinormal 
modes, we might need to wait for the LISA \cite{2022_gogoi}. The reason behind 
this is that assuming the nearest black holes, such as Sgr A$^*$ or M $87^*$ 
to be almost static, their respective quasinormal modes are beyond the 
sensitivity range of current and near future ground-based detectors 
\cite{2022_gogoi}. Nonetheless, the upcoming LISA space-based detector 
exhibits a remarkable increase in sensitivity and will be proficient in 
detecting quasinormal modes with a greater precision.

We also study the optical properties of the black hole {\it viz.,} the shadow 
and the emission rate. It is seen from our investigation that the shadow of 
the black hole decreases with an increase in the value of the parameter $c_0$. 
Similarly, an increase in the value of $c_0$ decreases the emission 
rate or evaporation rate of the black hole, hence increasing its stability.

Our investigation demonstrates that the boundary-derived parameter $c_0$ 
exerts considerable influence on the spacetime of a black hole, impacting 
various observable characteristics of it. We demonstrate that $c_0$ has 
effects on the gravitational deflection of light, ringdown GWs or quasinormal 
modes, the black hole shadow and the emission rate, potentially providing 
observable evidence for $f(\mathcal{T}, \mathcal{B})$ gravity.  Constraining 
the model with available observational data will provide more useful insights 
into the $f(\mathcal{T}, \mathcal{B})$ gravity which we keep as a future 
prospect of our study.

\section*{Acknowledgments}
UDG is thankful to the Inter-University Centre for Astronomy and Astrophysics
(IUCAA), Pune, India for the Visiting Associateship of the institute.

\bibliographystyle{apsrev}
\end{document}